\documentclass[journal]{IEEEtran} \newcommand{\figsize}{\linewidth}
\usepackage{graphicx,psfrag}
\usepackage[cmex10]{amsmath}
\usepackage{amsfonts,amssymb,latexsym,hyperref}
\usepackage[nospace,nocompress]{cite}
\usepackage[usenames,dvipsnames]{color}
\usepackage{subfig}
\usepackage{algorithm,algorithmic,setspace}
\usepackage{xpatch} 
\usepackage{multirow}

\makeatletter
\def\BState{\State\hskip-\ALG@thistlm}
\xpatchcmd{\algorithmic}{\itemsep\z@}{\itemsep=0.5ex plus2pt}{}{} 
\makeatother

 \newcommand{\mycaptionfont}{small}

 \renewcommand{\tilde}{\widetilde}
 \renewcommand{\hat}{\widehat}

 \newcommand{\defn}{\triangleq}

 \newcommand{\tvec}[1]{\ensuremath{\Tilde{\boldsymbol{#1}}}}
 
 \newcommand{\hvec}[1]{\ensuremath{\Hat{\boldsymbol{#1}}}}
    
 \renewcommand{\vec}[1]{\ensuremath{\boldsymbol{#1}}}
 \newcommand{\mat}[1]{\ensuremath{\begin{bmatrix}#1\end{bmatrix}}}

 \newcommand{\norm}[1]{\ensuremath{\| #1 \|}}
 \newcommand{\mc}[1]{\ensuremath{\mathcal{#1}}}
 \newcommand{\st}{{~\text{s.t.}~}}
 \newcommand{\barst}[1]{\ensuremath{\text{\raisebox{-0.5mm}{$\bigl|_{#1}$}}}}

 \newcommand{\Real}{{\mathbb{R}}}

 \newcommand{\floor}[1]{\left\lfloor #1 \right\rfloor}
 
 \newcommand{\of}[1]{^{(#1)}}

 \newcommand{\tran}{^{\textsf{T}}}
 
 \newcommand{\giv}{\,|\,}
 \newcommand{\biggiv}{\,\big|\,}
 
 \newcommand{\normal}{\mathcal{N}} 
 \newcommand*\deriv{\mathop{}\!\mathrm{d}}
 \newcommand{\tm}{t\!-\!}
 \newcommand{\inv}{^{-1}}

 \DeclareMathOperator{\sgn}{sgn}
 
 \DeclareMathOperator{\E}{E}
 \DeclareMathOperator{\var}{var}
 \DeclareMathOperator{\cov}{Cov}

 \DeclareMathOperator{\Diag}{Diag}

 \DeclareMathOperator*{\argmin}{arg\,min}
 \DeclareMathOperator*{\argmax}{arg\,max}

 \renewcommand{\eqref}[1]{(\ref{eq:#1})}
 
 \newcommand{\Figref}[1]{Figure~\ref{fig:#1}}
 
 \newcommand{\figref}[1]{Fig.~\ref{fig:#1}}
 
 \newcommand{\tabref}[1]{Table~\ref{tab:#1}}
 \newcommand{\secref}[1]{Section~\ref{sec:#1}}

 \newcommand{\algref}[1]{Algorithm~\ref{alg:#1}}
 \newcommand{\lineref}[1]{line~\ref{line:#1}}
 \newcommand{\Lineref}[1]{Line~\ref{line:#1}}

 \newcommand{\textb}[1]{\textcolor{black}{#1}}
 \newcommand{\blue}{black}

 \newfloat{myalgo}{tbhp}{mya}
  {\begin{myalgo}[#1]
        \centering
        \begin{minipage}{#2}
        \begin{algorithm}[H]}
  {\end{algorithm}
        \end{minipage}
        \end{myalgo}}

\newcommand{\rveca}{\textsf{\textbf{a}}}

\newcommand{\ry}{\textsf{y}}
\newcommand{\rvecy}{\textsf{\textbf{y}}}

\newcommand{\rvecz}{\textsf{\textbf{z}}}

\newcommand{\rvecp}{\textsf{\textbf{p}}}
\newcommand{\rvecs}{\textsf{\textbf{s}}}
\newcommand{\Qz}{\vec{Q}^{\rvecz}}
\newcommand{\Qp}{\vec{Q}^{\rvecp}}
\newcommand{\Qs}{\vec{Q}^{\rvecs}}
\newcommand{\qz}{q^{\rvecz}}
\newcommand{\qp}{q^{\rvecp}}
\newcommand{\qs}{q^{\rvecs}}
\newcommand{\rx}{\textsf{x}}
\newcommand{\rr}{\textsf{r}}

\newcommand{\rvecx}{\textbf{\textsf{x}}}
\newcommand{\rvecX}{\textbf{\textsf{X}}}
\newcommand{\rvecr}{\textbf{\textsf{r}}}
\newcommand{\rvecv}{\textbf{\textsf{v}}}
\newcommand{\Qr}{\vec{Q}^{\rvecr}}
\newcommand{\Qx}{\vec{Q}^{\rvecx}}
\newcommand{\qr}{q^{\rvecr}}
\newcommand{\qx}{q^{\rvecx}}

\newcommand{\pyz}{p_{\ry|\rvecz}}
\newcommand{\px}{p_{\rvecx}}
\newcommand{\pxscalar}{p_{\rx}}
\newcommand{\pxr}{p_{\rvecx|\rvecr}}
\newcommand{\pxrscalar}{p_{\rx|\rr}}
\newcommand{\pzyp}{p_{\rvecz|\ry,\rvecp}}
\newcommand{\pr}{p_{\textsf{r}}(r)}
\newcommand{\appropto}{\mathrel{\vcenter{
  \offinterlineskip\halign{\hfil$##$\cr
    \propto\cr\noalign{\kern2pt}\sim\cr\noalign{\kern-2pt}}}}}

\newcommand{\Mtraint}{M_{\textsf{train},t}}
\newcommand{\Mtestt}{M_{\textsf{test},t}}
\newcommand{\Mtest}{M_{\textsf{test}}}

\begin{document}
\setlength{\arraycolsep}{0.8mm}
 \title{Sparse Multinomial Logistic Regression via Approximate Message Passing}
 \author{Evan Byrne and Philip Schniter\IEEEauthorrefmark{1}%
        \thanks{The authors are with the 
		Department of Electrical and Computer Engineering
                at The Ohio State University, Columbus, OH.}%
        \thanks{Please direct all correspondence to 
		Prof.\ Philip Schniter,
                Dept. ECE, 2015 Neil Ave., Columbus OH 43210,
                e-mail: schniter@ece.osu.edu,
                phone 614.247.6488, fax 614.292.7596.
		Evan Byrne can be reached at the same address/phone/fax 
		and e-mailed at byrne.133@osu.edu.}%
    	\thanks{This work was supported in part by 
                the National Science Foundation grants CCF-1018368
                and CCF-1218754.}%
        \thanks{Portions of this work were presented at the 
                2015 Duke Workshop on Sensing and Analysis of High Dimensional Data.}%
        }
 \date{\today}
 \maketitle

\begin{abstract}
For the problem of multi-class linear classification and feature selection, we propose approximate message passing approaches to sparse multinomial logistic regression (MLR).
First, we propose two algorithms based on the Hybrid Generalized Approximate Message Passing (HyGAMP) framework:
one finds the maximum a posteriori (MAP) linear classifier and the other finds an approximation of the test-error-rate minimizing linear classifier.
Then we design computationally simplified variants of these two algorithms. 
Next, we detail methods to tune the hyperparameters of their assumed statistical models using Stein's unbiased risk estimate (SURE) and expectation-maximization (EM), respectively.
Finally, using both synthetic and real-world datasets, we demonstrate improved error-rate and runtime performance relative to \textb{existing state-of-the-art approaches to sparse MLR}.
\end{abstract}


\section{Introduction}              \label{sec:intro}

\subsection{Objective} \label{sec:objective}
We consider the problems of multiclass (or polytomous) linear classification and feature selection.
In both problems, one is given training data of the form $\{(y_m,\vec{a}_m)\}_{m=1}^M$, where $\vec{a}_m\in\Real^N$ is a vector of features and $y_m\in\{1,\dots,D\}$ is the corresponding $D$-ary class label. 
In \emph{multiclass classification}, the goal is to infer the unknown label $y_0$ associated with a newly observed feature vector $\vec{a}_0$.
In the \emph{linear} approach to this problem, the training data \textb{are} used to design a weight matrix $\vec{X}\in\Real^{N\times D}$ that generates a vector of ``scores'' $\vec{z}_0 \defn \vec{X}\tran\vec{a}_0 \in\Real^D$, the largest of which can be used to predict the unknown label, i.e., 
\begin{align}
\hat{y}_0 
&= \arg\max_d \, [\vec{z}_0]_d.
\label{eq:predict}
\end{align}
In \emph{feature selection}, the goal is to determine which \emph{subset} of the $N$ features $\vec{a}_0$ is needed to accurately predict the label $y_0$.

We are particularly interested in the setting where the number of features, $N$, is large and greatly exceeds the number of training examples, $M$.
Such problems arise in a number of important applications, such as 
micro-array gene expression \cite{Sun:CAN:06,Bhattacharjee:PNAS:01}, 
multi-voxel pattern analysis (MVPA) \cite{Haxby:SCI:01,NPDH2006}, 
text mining \cite{F2003b,Lewis:MLR:04}, 
and analysis of marketing data \cite{GHH2007}. 


%
In the $N\gg M$ case, accurate linear classification and feature selection may be possible if the labels are influenced by a sufficiently small number, $K$, of the total $N$ features.
For example, in binary linear classification, performance guarantees are possible with only $M=O(K\log N/K)$ training examples when $\vec{a}_m$ is i.i.d.\ Gaussian \cite{Plan:TIT:13}.
Note that, when $K\ll N$, accurate linear classification can be accomplished using a \emph{sparse} weight matrix $\vec{X}$, i.e., a matrix where all but a few rows are zero-valued.

\subsection{Multinomial logistic regression} \label{sec:MLR}

For multiclass linear classification and feature selection, we focus on the approach known as \emph{multinomial logistic regression} (MLR) \cite{Bishop:Book:07}, which can be described using a generative probabilistic model.
Here, the label vector $\vec{y}\defn[y_0,\dots,y_M]\tran$ is modeled as a realization of a random\footnote{For clarity, we typeset random quantities in sans-serif font and deterministic quantities in serif font.} vector $\rvecy\defn[\ry_0,\dots,\ry_M]\tran$, the ``true'' weight matrix $\vec{X}$ is modeled as a realization of a random matrix $\rvecX$, and the features $\vec{A}\defn[\vec{a}_0,\dots,\vec{a}_M]\tran$ are treated as deterministic. 
Moreover, the labels $\ry_m$ are modeled as conditionally independent given the scores $\rvecz_m\defn\rvecX\tran\vec{a}_m$, i.e.,
\begin{align}
\Pr\{\rvecy=\vec{y} \giv \rvecX=\vec{X};\vec{A}\} 
&= \prod_{m=1}^M \pyz(y_m|\vec{X}\tran\vec{a}_m) ,
\label{eq:py|X}
\end{align}
and distributed according to the multinomial logistic (or soft-max) pmf: 
\begin{align}
\pyz(y_m|\vec{z}_m)
&= \frac{\exp([\vec{z}_m]_{y_m})}{\sum_{d=1}^D \exp([\vec{z}_m]_{d})},
~~y_m\in\{1,\dots,D\} .
\label{eq:softmax}
\end{align}
The rows $\rvecx_n\tran$ of the weight matrix $\rvecX$ are then modeled as i.i.d., 
\begin{align}
p_{\rvecX}(\vec{X})
&= \prod_{n=1}^N \px(\vec{x}_n) ,
\label{eq:pX}
\end{align}
where $\px$ may be chosen to promote sparsity.

\subsection{Existing methods} \label{sec:existing}

Several \textb{sparsity-promoting} MLR algorithms have been proposed
(e.g., \cite{Tipping:JMLR:01,Krishnapuram:TPAMI:05,Genkin:Techno:07,Friedman:JSS:10,Cawley:NIPS:07,Meier:JRSSb:08}), differing in their choice of $\px$ and methodology of estimating $\rvecX$.
For example, 
\cite{Krishnapuram:TPAMI:05,Genkin:Techno:07,Friedman:JSS:10} use the i.i.d.\ Laplacian prior
\begin{align}
\px(\vec{x}_n;\lambda)
&= \prod_{d=1}^D \frac{\lambda}{2} \exp(-\lambda|x_{nd}|) ,
\label{eq:laplace}
\end{align}
with $\lambda$ tuned via cross-validation.
To circumvent this tuning problem,
\cite{Cawley:NIPS:07} employs the Laplacian scale mixture
\begin{align}
\px(\vec{x}_n)
&= \prod_{d=1}^D \int \left[\frac{\lambda}{2} \exp(-\lambda|x_{nd}|) \right] 
        p(\lambda) \deriv\lambda ,
\label{eq:lsm}
\end{align}
with Jeffrey's non-informative hyperprior $p(\lambda)\propto\frac{1}{\lambda} 1_{\lambda\geq 0}$. 
The relevance vector machine (RVM) approach \cite{Tipping:JMLR:01} uses the Gaussian scale mixture 
\begin{align}
\px(\vec{x}_n)
&= \prod_{d=1}^D \int \normal(x_{nd};0,\nu)
        p(\nu) \deriv\nu ,
\label{eq:gsm}
\end{align}
with inverse-gamma $p(\nu)$ (i.e., the conjugate hyperprior), resulting in an i.i.d.\ student's t distribution for $\px$. 
However, other choices are possible.
For example, the exponential hyperprior $p(\nu;\lambda)=\frac{\lambda^2}{2}\exp(-\frac{\lambda^2}{2}\nu) 1_{\nu\geq 0}$ would lead back to the i.i.d.\ Laplacian distribution \eqref{laplace} for $\px$ \cite{Grandvalet:ICANN:98}.
Finally, 
\cite{Meier:JRSSb:08} uses 
\begin{align}
\px(\vec{x}_n;\lambda)\propto \exp(-\lambda \|\vec{x}_n\|_2) ,
\label{eq:group}
\end{align}
which encourages row-sparsity in $\vec{X}$.

Once the probabilistic model \eqref{py|X}-\eqref{pX} has been specified, a procedure is needed to infer the weights $\rvecX$ from the training data $\{(y_m,\vec{a}_m)\}_{m=1}^M$.
The Laplacian-prior methods \cite{Krishnapuram:TPAMI:05,Genkin:Techno:07,Friedman:JSS:10,Meier:JRSSb:08} use the maximum a posteriori (MAP) estimation framework:
\begin{align}
\lefteqn{ \hvec{X}
= \arg\max_{\vec{X}} \log p(\vec{X}|\vec{y};\vec{A}) }\\
&= \arg\max_{\vec{X}} \sum_{m=1}^M \log \pyz(y_m|\vec{X}\tran\vec{a}_m) + \sum_{n=1}^N \log \px(\vec{x}_n) , 
\label{eq:MAP}
\end{align}
where Bayes' rule was used for \eqref{MAP}.
Under $\px$ from \eqref{laplace} or \eqref{group}, the second term in \eqref{MAP} reduces to 
$-\lambda\sum_{n=1}^N \|\vec{x}_n\|_1$ or 
$-\lambda\sum_{n=1}^N \|\vec{x}_n\|_2$, respectively. 
In this case, \eqref{MAP} is concave and can be maximized in polynomial time; \cite{Krishnapuram:TPAMI:05,Genkin:Techno:07,Friedman:JSS:10,Meier:JRSSb:08} employ (block) coordinate ascent for this purpose.
The papers \cite{Tipping:JMLR:01} and \cite{Cawley:NIPS:07} handle the scale-mixture priors \eqref{lsm} and \eqref{gsm}, respectively, using the evidence maximization framework \cite{Mackay:NC:92b}. 
This approach yields a double-loop procedure: the hyperparameter $\lambda$ or $\nu$ is estimated in the outer loop, and---for fixed $\lambda$ or $\nu$---the resulting concave (i.e., $\ell_2$ or $\ell_1$ regularized) MAP optimization is solved in the inner loop.

The methods \cite{Tipping:JMLR:01,Krishnapuram:TPAMI:05,Genkin:Techno:07,Friedman:JSS:10,Cawley:NIPS:07,Meier:JRSSb:08} described above all yield a sparse point estimate $\hvec{X}$. 
Thus, feature selection is accomplished by examining the row-support of $\hvec{X}$ and classification is accomplished through \eqref{predict}.

\subsection{Contributions} \label{sec:contribution}

In \secref{hygamp},
we propose new approaches to sparse-weight MLR based on the \emph{hybrid generalized approximate message passing} (HyGAMP) framework from \cite{Rangan:ISIT:12}. 
HyGAMP offers tractable approximations of the sum-product and min-sum message passing algorithms \cite{Pearl:Book:88} by leveraging results of the central limit theorem that hold in the large-system limit: $\lim_{N,M\rightarrow\infty}$ with fixed $N/M$.
Without approximation, both the sum-product algorithm (SPA) and min-sum algorithm (MSA) are intractable due to the forms of $\pyz$ and $\px$ in our problem.

For context, we note that HyGAMP is a generalization of the original GAMP approach from \cite{Rangan:ISIT:11}, which cannot be directly applied to the MLR problem because the likelihood function \eqref{softmax} is not separable, i.e., $\pyz(y_m|\vec{z}_m)\neq\prod_d p(y_m|z_{md})$.
GAMP can, however, be applied to \emph{binary} classification and feature selection, as in \cite{Ziniel:TSP:15}. 
Meanwhile, GAMP is itself a generalization of the original AMP approach from \cite{Donoho:PNAS:09,Donoho:ITW:10a}, which requires $\pyz$ to be both separable and Gaussian.

With the HyGAMP algorithm from \cite{Rangan:ISIT:12}, message passing for sparse-weight MLR reduces to an iterative update of $O(M+N)$ multivariate Gaussian pdfs, each of dimension $D$.
Although HyGAMP makes MLR tractable, it is still not computationally practical for the large values of $M$ and $N$ in contemporary applications (e.g., $N\sim 10^4$ \textb{to $10^6$} in genomics and MVPA).
Similarly, the non-conjugate variational message passing technique from \cite{Knowles:NIPS:11} requires the update of $O(MN)$ multivariate Gaussian pdfs of dimension $D$, which is even less practical for large $M$ and $N$. 

Thus, in \secref{shygamp}, we propose a simplified HyGAMP (SHyGAMP) algorithm for MLR that approximates HyGAMP's mean and variance computations in an efficient manner.
In particular, we investigate approaches based on 
numerical integration, 
importance sampling,
Taylor-series approximation, 
and a novel Gaussian-mixture approximation,
and we conduct numerical experiments that suggest the superiority of the latter.

In \secref{param_tune}, we detail two approaches to tune the hyperparameters that control the statistical models assumed by SHyGAMP, one based on the expectation-maximization (EM) methodology from \cite{Vila:TSP:13} and the other based on a variation of the Stein's unbiased risk estimate (SURE) methodology from \cite{Mousavi:13}.
We also give numerical evidence that these methods yield near-optimal hyperparameter estimates.

Finally, in \secref{sims}, we compare our proposed SHyGAMP methods to the state-of-the-art MLR approaches \cite{Friedman:JSS:10,Cawley:NIPS:07} on both synthetic and practical real-world problems.
Our experiments suggest that our proposed methods offer simultaneous improvements in classification error rate and runtime.



\emph{Notation:}
Random quantities are typeset in sans-serif (e.g., $\rx$) while deterministic quantities are typeset in serif (e.g., $x$).  
The pdf of random variable $\rx$ under deterministic parameters $\vec{\theta}$ is written as $p_{\rx}(x;\vec{\theta})$, where the subscript and parameterization are sometimes omitted for brevity.
Column vectors are typeset in boldface lower-case (e.g., $\vec{y}$ or $\rvecy$), matrices in boldface upper-case (e.g., $\vec{X}$ or $\rvecX$), and their transpose is denoted by $(\cdot)\tran$.
 $\E\{\cdot\}$ denotes expectation
 and $\cov\{\cdot\}$ autocovariance.
 %
 $\vec{I}_K$ denotes the $K \times K$ identity matrix,
 $\vec{e}_k$ the $k$th column of $\vec{I}_K$,
 $\vec{1}_K$ the length-$K$ vector of ones,
 and
 $\Diag(\vec{b})$ the diagonal matrix created from the vector $\vec{b}$.
 $[\vec{B}]_{m,n}$ denotes the element in the $m^{th}$ row and $n^{th}$ column of $\vec{B}$, 
 %
 and
 $\norm{\cdot}_F$ the Frobenius norm. 
 %
 Finally,
 $\delta_n$ denotes the Kronecker delta sequence,
 $\delta(x)$ the Dirac delta distribution,
 and 
 $1_{A}$ the indicator function of the event $A$.

\section{HyGAMP for Multiclass Classification} \label{sec:hygamp}

In this section, we detail the application of HyGAMP \cite{Rangan:ISIT:12} to multiclass linear classification.
In particular, we show that the sum-product algorithm (SPA) variant of HyGAMP is a loopy belief propagation (LBP) approximation of the classification-error-rate minimizing linear classifier and that the min-sum algorithm (MSA) variant is an LBP approach to solving the MAP problem \eqref{MAP}.

\subsection{Classification via sum-product HyGAMP} \label{sec:spa_hygamp}

\begin{figure}
\centering
\subfloat[][Full graph]{
 \newcommand{\bs}{1.0}
 \psfrag{pyz}[bl][bl][\bs]{$p_{{\ry_m|\rvecz_m}}$}
 \psfrag{plz}[bl][bl][\bs]{$p_{{\ry_t|\rvecz_t}}$}
 \psfrag{x}[bl][Bl][\bs]{$\rvecx_n$}
 \psfrag{y}[bl][Bl][\bs]{$\ry_m$}
 \psfrag{l}[bc][Bc][\bs]{$\ry_t$}
 \psfrag{pxh}[bl][Bl][\bs]{$p_{\rvecx_n}$}
 \scalebox{0.75}{\includegraphics{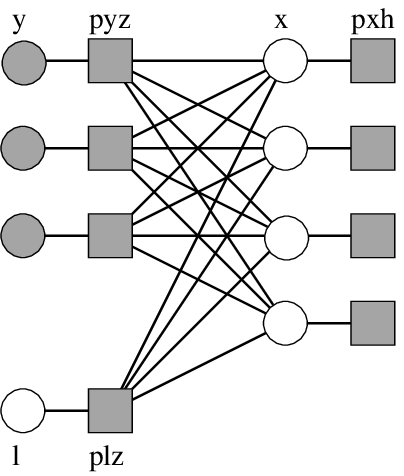}}
\label{fig:full_factor_graph}
}
\quad
\subfloat[][Reduced graph]{
 \newcommand{\bs}{1.0}
 \psfrag{pyz}[bl][bl][\bs]{$p_{{\ry_m|\rvecz_m}}$}
 \psfrag{plz}[bl][bl][\bs]{$p_{{\ry_t|\rvecz_t}}$}
 \psfrag{x}[bl][Bl][\bs]{$\rvecx_n$}
 \psfrag{y}[bl][Bl][\bs]{$\ry_m$}
 \psfrag{l}[bc][Bc][\bs]{$\ry_t$}
 \psfrag{pxh}[bl][Bl][\bs]{$p_{\rvecx_n}$}
 \scalebox{0.75}{\includegraphics{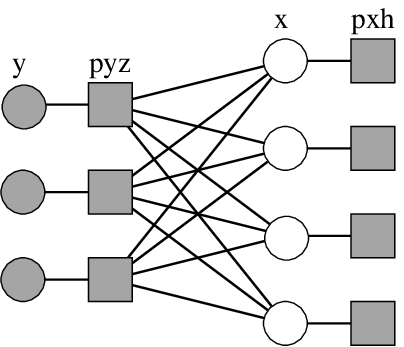}}
\label{fig:reduced_factor_graph}
}
\captionsetup{font=\mycaptionfont}
\caption{Factor graph representations of \eqref{predict_factor_rep}, with white/gray circles denoting unobserved/observed random variables, and gray rectangles denoting pdf ``factors''.}
\label{fig:discriminative_factor_graph}
\end{figure}

Suppose that we are given $M$ labeled training pairs $\{(y_m,\vec{a}_m)\}_{m=1}^M$ and $T$ test feature vectors $\{\vec{a}_t\}_{t=M+1}^{M+T}$ associated with unknown test labels $\{\ry_t\}_{t=M+1}^{M+T}$, all obeying the MLR statistical model \eqref{py|X}-\eqref{pX}.
Consider the problem of computing the classification-error-rate minimizing hypotheses $\{\hat{y}_t\}_{t=M+1}^{M+T}$,
\begin{align}
\hat{y}_t
&= \arg\max_{y_t\in\{1,\dots,D\}} 
p_{\ry_t|\rvecy_{1:M}}\big(y_t \biggiv \vec{y}_{1:M}; \vec{A}\big),
\label{eq:yhatt}
\end{align}
under known $\pyz$ and $\px$, 
where $\vec{y}_{1:M}\defn [y_1,\dots,y_M]\tran$ and $\vec{A}\defn[\vec{a}_1,\dots,\vec{a}_{M+T}]\tran$.
The probabilities in \eqref{yhatt} can be computed via the marginalization
\begin{align}
\lefteqn{ 
p_{\ry_t|\rvecy_{1:M}}\big(y_t\biggiv\vec{y}_{1:M};\vec{A}\big)
= p_{\ry_t,\rvecy_{1:M}}\big(y_t,\vec{y}_{1:M};\vec{A}\big) Z_{\rvecy}^{-1} 
}\\
&= Z_{\rvecy}^{-1} \sum_{\vec{y}\in\mathcal{Y}_t(y_t)} 
        \int p_{\rvecy,\rvecX}(\vec{y},\vec{X};\vec{A}) \deriv\vec{X}, \hspace{0.75in}
\label{eq:marginalize_labels}
\end{align}
with scaling constant $Z_{\rvecy}^{-1}$, 
label vector $\vec{y}=[y_1,\dots,y_{M+T}]\tran$, 
and constraint set $\mathcal{Y}_t(y)\defn \big\{\tvec{y}\in\{1,\dots,D\}^{M+T}\st [\tvec{y}]_t=y \text{~and~}[\tvec{y}]_m=y_m~\forall m=1,\dots,M \big\}$,
which fixes the $t$th element of $\vec{y}$ at the value $y$ and the first $M$ elements of $\vec{y}$ at the values of the corresponding training labels.
Due to \eqref{py|X} and \eqref{pX}, the joint pdf in \eqref{marginalize_labels} factors as
\begin{eqnarray}
p_{\rvecy,\rvecX}(\vec{y},\vec{X};\vec{A})
&=& \prod_{m=1}^{M+T} p_{\ry|\rvecz}(y_m\giv \vec{X}\tran\vec{a}_m)\, \prod_{n=1}^{N}p_{\rvecx}(\vec{x}_n) .
\qquad
\label{eq:predict_factor_rep} 
\end{eqnarray}
The factorization in \eqref{predict_factor_rep} is depicted by the \emph{factor graph} in \figref{full_factor_graph}, where the random variables $\{\ry_m\}$ and random vectors $\{\rvecx_n\}$ are connected to the pdf factors in which they appear.

Since exact computation of the marginal posterior test-label probabilities is an NP-hard problem \cite{Cooper:AI:90}, we are interested in alternative strategies, such as those based on loopy belief propagation by the SPA \cite{Pearl:Book:88}.
Although a direct application of the SPA is itself intractable when $\pyz$ takes the MLR form \eqref{softmax}, the SPA simplifies in the large-system limit under i.i.d.\ sub-Gaussian $\vec{A}$, leading to the HyGAMP approximation \cite{Rangan:ISIT:12} given\footnote{%
The HyGAMP algorithm in \cite{Rangan:ISIT:12} is actually more general than what is specified in \algref{hygamp}, but the version in \algref{hygamp} is sufficient to handle the factor graph in \figref{full_factor_graph}.}
in \algref{hygamp}.
Although in practical MLR applications $\vec{A}$ is not i.i.d.\ Gaussian,\footnote{\textb{We note that many of the standard data pre-processing techniques, such as z-scoring, tend to make the feature distributions closer to zero-mean Gaussian.}} the numerical results in \secref{sims} suggest that treating it as such works sufficiently well.

We note from \figref{full_factor_graph} that the HyGAMP algorithm is applicable to a factor graph with vector-valued variable nodes.
As such, it generalizes the GAMP algorithm from \cite{Rangan:ISIT:11}, which applies only to a factor graph with scalar-variable nodes. 
Below, we give a brief explanation for the steps in \algref{hygamp}.
For those interested in more details, we suggest
\cite{Rangan:ISIT:12} for an overview and derivation of HyGAMP,
\cite{Rangan:ISIT:11} for an overview and derivation of GAMP,
\cite{Javanmard:II:13} for rigorous analysis of GAMP under large i.i.d.\ sub-Gaussian $\vec{A}$, and 
\cite{Rangan:ISIT:13,Rangan:ISIT:14} for fixed-point and local-convergence analysis of GAMP under arbitrary $\vec{A}$.

Lines~\ref{line:spaxhat}-\ref{line:spaQx} of \algref{hygamp} produce an approximation of the posterior mean and covariance of $\rvecx_n$ at each iteration $t$.
Similarly, lines~\ref{line:spazhat}-\ref{line:spaQz} produce an approximation of the posterior mean and covariance of $\rvecz_m\defn \rvecX\tran \vec{a}_m$.
The posterior mean and covariance of $\rvecx_n$ are computed from the intermediate quantity $\hvec{r}_n(t)$, which behaves like a noisy measurement of the true $\vec{x}_n$.
In particular, for i.i.d.\ Gaussian $\vec{A}$ in the large-system limit, $\hvec{r}_n(t)$ is a typical realization of the random vector $\rvecr_n=\vec{x}_n+\rvecv_n$ with $\rvecv_n\sim\normal(\vec{0},\Qr_n(t))$.
Thus, the approximate posterior pdf used in lines~\ref{line:spaxhat}-\ref{line:spaQx} is
\begin{align} 
\pxr(\vec{x}_n|\hvec{r}_n;\Qr_n) 
&= 
\frac{\px(\vec{x}_n)\normal(\vec{x}_n;\hvec{r}_n,\Qr_n)}
        {\int \px(\vec{x}'_n)\normal(\vec{x}'_n;\hvec{r}_n,\Qr_n) \deriv \vec{x}'_n} 
\label{eq:pxr}.
\end{align} 
A similar interpretation holds for HyGAMP's approximation of the posterior mean and covariance of $\rvecz_m$ in lines~\ref{line:spazhat}-\ref{line:spaQz}, which uses the intermediate vector $\hvec{p}_m(t)$ and the approximate posterior pdf 
\begin{align} 
\lefteqn{ \pzyp(\vec{z}_m|y_m,\hvec{p}_m;\Qp_m) }\nonumber\\
&= 
\frac{\pyz(y_m|\vec{z}_m)\normal(\vec{z}_m;\hvec{p}_m,\Qp_m)}
        {\int \pyz(y_m|\vec{z}'_m)\normal(\vec{z}'_m;\hvec{p}_m,\Qp_m) \deriv \vec{z}'_m} .
\label{eq:pzyp}
\end{align} 

\begin{algorithm}[t]
\footnotesize
\caption{HyGAMP}
\label{alg:hygamp}
\begin{algorithmic}[1]
\REQUIRE{%
        Mode $\in\{\texttt{SPA},\texttt{MSA}\}$,
        matrix $\vec{A}$, vector $\vec{y}$,
        pdfs $\pxr$ and $\pzyp$ from \eqref{pxr}-\eqref{pzyp},
        initializations 
                $\hvec{r}_n(0)$,
                $\Qr_n(0)$.
        }
\ENSURE{$t \!\leftarrow\! 0$;\,
        $\hvec{s}_m(0) \!\leftarrow\! \vec{0}$.
        }
\REPEAT 
        \IF[$\textbf{for}~n=1\dots N$]{\texttt{MSA}}
        	\STATE 
                $\hvec{x}_n(t) \leftarrow \argmax_{\vec{x}} \log \pxr\big(\vec{x}_n \big|
                                \hvec{r}_n(\tm1) ; \Qr_n(\tm1) \big)$ 
                        \label{line:msaxhat}
                \STATE 
                $\Qx_n(t) \leftarrow \big[-\frac{\partial^2}{\partial \vec{x}^2}  
                        \log \pxr\big(\hvec{x}_n(t) \big| 
                                \hvec{r}_n(\tm1) ; \Qr_n(\tm1) \big) \big]^{-1}$
                        \label{line:msaQx}
        \ELSIF[$\textbf{for}~n=1\dots N$]{\texttt{SPA}}
        \STATE  
                $\hvec{x}_n(t) \leftarrow \E\big\{\rvecx_n \biggiv \rvecr_n = \hvec{r}_n(\tm1); \Qr_n(\tm1)\big\}$ 
                        \label{line:spaxhat}
                \STATE 
                $\Qx_n(t) \leftarrow \cov\big\{\rvecx_n \biggiv \rvecr_n = \hvec{r}_n(\tm1); \Qr_n(\tm1)\big\}$ 
                        \label{line:spaQx}
        \ENDIF
        \STATE 
        $\forall m:\, \Qp_m (t) \leftarrow \sum_{n=1}^N {A}_{mn}^2 \Qx_n(t)$ 
                        \label{line:Qp}
        \STATE 
        $\forall m:\, \hvec{p}_m (t) \leftarrow \sum_{n=1}^N A_{mn} \hvec{x}_n(t) - \Qp_m(t) \hvec{s}_m(\tm1)$ 
                        \label{line:phat}
        \IF[$\textbf{for}~m=1\dots M$]{\texttt{MSA}}
        \STATE 
                $\hvec{z}_m(t) \leftarrow \argmax_{\vec{z}} \log \pzyp\big(\vec{z}_m \big| y_m,
                                \hvec{p}_m(t) ; \Qp_m(t) \big)$ 
                        \label{line:msazhat}
                \STATE 
                $\Qz_m(t) \leftarrow \big[-\frac{\partial^2}{\partial \vec{z}^2}  
                        \log \pzyp\big(\hvec{z}_m(t) \big| y_m,
                                \hvec{p}_m(t) ; \Qp_m(t) \big) \big]^{-1}$
                        \label{line:msaQz} 
        \ELSIF[$\textbf{for}~m=1\dots M$]{\texttt{SPA}}
                        \STATE  
                $\hvec{z}_m(t) \leftarrow \E\big\{\rvecz_m \biggiv y_m, \rvecp_m = \hvec{p}_m(t); \Qp_m(t)\big\}$ 
                        \label{line:spazhat}
                \STATE 
                $\Qz_m(t) \leftarrow \cov\big\{\rvecz_m \biggiv y_m, \rvecp_m = \hvec{p}_m(t); \Qp_m(t)\big\}$ 
                        \label{line:spaQz}
        \ENDIF
        \STATE 
        $\forall m:\, \Qs_m(t) \leftarrow [\Qp_m(t)]^{-1} - [\Qp_m(t)]^{-1} \Qz_m (t) [\Qp_m(t)]^{-1}$ 
                        \label{line:Qs}
        \STATE 
        $\forall m:\, \hvec{s}_m(t) \leftarrow [\Qp_m(t)]^{-1} \big(\hvec{z}_m(t) - \hvec{p}_m(t) \big)$ 
                        \label{line:shat}
        \STATE 
        $\forall n:\, \Qr_n(t) \leftarrow \big[\sum_{m=1}^M A_{mn}^2 \Qs_m(t) \big]^{-1}$
                        \label{line:Qr}
        \STATE 
        $\forall n:\, \hvec{r}_n(t) \leftarrow \hvec{x}_n(t) + \Qr_n(t) \sum_{m=1}^M A_{mn} \hvec{s}_m(t)$ 
                        \label{line:rhat}
        \STATE 
        $t \leftarrow t + 1$
\UNTIL{Terminated}
\end{algorithmic}
\end{algorithm}

\subsection{Classification via min-sum HyGAMP} \label{sec:msa_hygamp}

As discussed in \secref{existing}, an alternative approach to linear classification and feature selection is through MAP estimation of the true weight matrix $\vec{X}$.
Given a likelihood of the form \eqref{py|X} and a prior of the form \eqref{pX}, the MAP estimate is the solution to the optimization problem \eqref{MAP}.

Similar to how the SPA can be used to compute approximate marginal posteriors in loopy graphs, the min-sum algorithm (MSA) \cite{Pearl:Book:88} can be used to compute the MAP estimate.
Although a direct application of the MSA is intractable when $\pyz$ takes the MLR form \eqref{softmax}, the MSA simplifies in the large-system limit under i.i.d.\ sub-Gaussian $\vec{A}$, leading to the \texttt{MSA} form of HyGAMP specified in \algref{hygamp}.

As described in \secref{spa_hygamp}, when $\vec{A}$ is large and i.i.d.\ sub-Gaussian, the vector $\hvec{r}_n(t)$ in \algref{hygamp} behaves like a Gaussian-noise-corrupted observation of the true $\vec{x}_n$ with noise covariance $\Qr_n(t)$.
Thus, \lineref{msaxhat} can be interpreted as MAP estimation of $\vec{x}_n$ and \lineref{msaQx} as measuring the local curvature of the corresponding MAP cost. 
Similar interpretations hold for MAP estimation of $\vec{z}_m$ via lines~\ref{line:msazhat}-\ref{line:msaQz}.

\subsection{Implementation of sum-product HyGAMP} \label{sec:spa_hygamp_imp}

From \algref{hygamp}, we see that HyGAMP requires inverting $M+N$ matrices of size $D\times D$ (for lines~\ref{line:Qs} and \ref{line:Qr}) in addition to solving $M+N$ joint inference problems of dimension $D$ in lines~\ref{line:msaxhat}-\ref{line:spaQx} and \ref{line:msazhat}-\ref{line:spaQz}.
We now briefly discuss the latter problems for the sum-product version of HyGAMP.

\subsubsection{Inference of \texorpdfstring{$\vec{x}_n$}{x}} \label{sec:spa_hygamp_input}

One choice of weight-coefficient prior $p_{\rvecx_n}$ that facilitates row-sparse $\vec{X}$ and tractable SPA inference is Bernoulli-multivariate-Gaussian, i.e.,
\begin{align}
p_{\rvecx}(\vec{x}_n)
= (1-\beta)\delta(\vec{x}_n) + \beta \mc{N}(\vec{x}_n;\vec{0},v\vec{I}) ,
\label{eq:mBG}
\end{align}
where $\delta(\cdot)$ denotes the Dirac delta and $\beta\in(0,1]$.
In this case, it can be shown \cite{Byrne:Thesis:15} that the mean and variance computations in lines~\ref{line:spaxhat}-\ref{line:spaQx} of \algref{hygamp} reduce to 
\begin{align}
C_n
&= 1 + \frac{1-\beta}{\beta} \frac{ \normal(\vec{0}; \hvec{r}_n, \Qr_n)}{\normal(\vec{0}; \hvec{r}_n, v\vec{I} + \Qr_n)}\\
\hvec{x}_n
&= C_n^{-1} (\vec{I} + v^{-1}\Qr_n)\inv \hvec{r}_n \\
\Qx_n
&= C_n^{-1} (\vec{I} + v^{-1}\Qr_n)\inv \Qr_n
+ (C_n-1)  \hvec{x}_n \hvec{x}_n\tran 
,
\end{align}
which requires a $D\times D$ matrix inversion at each $n$.

\subsubsection{Inference of \texorpdfstring{$\vec{z}_m$}{z}} \label{sec:spa_hygamp_output}

When $p_{\ry|\rvecz}$ takes the MLR form in \eqref{softmax}, closed-form expressions for $\hvec{z}_m(t)$ and $\Qz_m(t)$ from lines~\ref{line:spazhat}-\ref{line:spaQz} of \algref{hygamp} do not exist.
While these computations could be approximated using, e.g., numerical integration or importance sampling, this is expensive because $\hvec{z}_m(t)$ and $\Qz_m(t)$ must be computed for every index $m$ at every HyGAMP iteration $t$. 
More details on these approaches will be presented in \secref{spa_shygamp_output}, in the context of SHyGAMP.

\subsection{Implementation of min-sum HyGAMP} \label{sec:msa_hygamp_imp}

\subsubsection{Inference of \texorpdfstring{$\vec{x}_n$}{x}} \label{sec:msa_hygamp_input}

To ease the computation of \lineref{msaxhat} in \algref{hygamp}, it is typical to choose a log-concave prior $p_{\rvecx}$ so that the optimization problem \eqref{MAP} is concave (since $p_{\ry|\rvecz}$ in \eqref{softmax} is also log-concave).
As discussed in \secref{existing}, a common example of a log-concave sparsity-promoting prior is the Laplace prior \eqref{laplace}.
In this case, \lineref{msaxhat} becomes
\begin{align}
\hvec{x}_n
&= \arg\max_{\vec{x}}
-\frac{1}{2}(\vec{x}-\hvec{r}_n)\tran[\Qr_n]^{-1} (\vec{x}-\hvec{r}_n) - \lambda \|\vec{x}\|_1 
\label{eq:lasso} ,
\end{align}
which is essentially the LASSO \cite{Tibshirani:JRSSb:96} problem.
Although \eqref{lasso} has no closed-form solution, it can be solved iteratively using, e.g., minorization-maximization (MM) \cite{Hunter:AS:04}.

To maximize a function $J(\vec{x})$, MM iterates the recursion
\begin{align}
\hvec{x}\of{k+1} 
&= \arg\max_{\vec{x}} \hat{J}(\vec{x};\hvec{x}\of{k}) 
\label{eq:MM} ,
\end{align}
where $\hat{J}(\vec{x};\hvec{x})$ is a surrogate function that minorizes $J(\vec{x})$ at $\hvec{x}$.
In other words, $\hat{J}(\vec{x};\hvec{x}) \leq J(\hvec{x}) ~\forall \vec{x}$ for any fixed $\hvec{x}$, with equality when $\vec{x}=\hvec{x}$.
To apply MM to \eqref{lasso}, we identify the utility function as
$J_n(\vec{x}) \defn -\frac{1}{2}(\vec{x}-\hvec{r}_n)\tran[\Qr_n]^{-1} (\vec{x}-\hvec{r}_n) - \lambda \|\vec{x}\|_1$.
Next we apply a result from \cite{Hunter:AS:05} that established that $J_n(\vec{x})$ is minorized by 
$\hat{J}_n(\vec{x};\hvec{x}\of{k}_n) \defn -\frac{1}{2}(\vec{x}-\hvec{r}_n)\tran[\Qr_n]^{-1} (\vec{x}-\hvec{r}_n) -\frac{\lambda}{2}\big(\vec{x}\tran \vec{\Lambda}(\hvec{x}\of{k}_n)\vec{x} + \|\hvec{x}\of{k}_n\|_2^2\big)$ 
with $\vec{\Lambda}(\hvec{x}) \defn\Diag\big\{|\hat{x}_1|^{-1},\dots,|\hat{x}_D|^{-1}\big\}$.
Thus \eqref{MM} implies 
\begin{align}
\lefteqn{
\hvec{x}_n\of{k+1} 
= \arg\max_{\vec{x}} \hat{J}_n(\vec{x};\hvec{x}\of{k}_n) 
}\\
&= \arg\max_{\vec{x}} \vec{x}\tran [\Qr_n]^{-1} \hvec{r}_n - \frac{1}{2}\vec{x}\tran \big( [\Qr_n]^{-1} + \lambda \vec{\Lambda}(\hvec{x}\of{k}_n) \big) \vec{x} 
\label{eq:mm1}\\
&= \big( [\Qr_n]^{-1} + \lambda \vec{\Lambda}(\hvec{x}\of{k}_n) \big)^{-1} [\Qr_n]^{-1} \hvec{r}_n
\label{eq:mm} 
\end{align}
where \eqref{mm1} dropped the $\vec{x}$-invariant terms from $\hat{J}_n(\vec{x};\hvec{x}\of{k}_n)$.
Note that each iteration $k$ of \eqref{mm} requires a $D\times D$ matrix inverse for each $n$.

\Lineref{msaQx} of \algref{hygamp} then says to set $\Qx_n$ equal to the Hessian of the objective function in \eqref{lasso} at $\hvec{x}_n$. 
Recalling that the second derivative of $|x_{nd}|$ is undefined when $x_{nd}=0$ but otherwise equals zero, we set
$\Qx_n=\Qr_n$ but then zero the $d$th row and column of $\Qx_n$ for all $d$ such that $\hat{x}_{nd}=0$.

\subsubsection{Inference of \texorpdfstring{$\vec{z}_m$}{z}} \label{sec:msa_hygamp_output}

Min-sum HyGAMP also requires the computation of lines~\ref{line:msazhat}-\ref{line:msaQz} in \algref{hygamp}. In our MLR application, \lineref{msazhat} reduces to the concave optimization problem
\begin{align}
\hvec{z}_m
&= \arg\max_{\vec{z}}
-\frac{1}{2}(\vec{z}-\hvec{p}_m)\tran[\Qp_m]^{-1} (\vec{z}-\hvec{p}_m) 
\nonumber\\&\quad 
+ \log \pyz(y_m|\vec{z}) 
\label{eq:mlr} .
\end{align}
Although \eqref{mlr} can be solved in a variety of ways
(see \cite{Byrne:Thesis:15} for MM-based methods),
we now describe one based on Newton's method \cite{Bertsekas:Book:99}, i.e., 
\begin{align}
\hvec{z}_m\of{k+1} 
&= \hvec{z}_m\of{k} - \alpha\of{k} [\vec{H}\of{k}_m]^{-1} \vec{g}\of{k}_m 
\label{eq:newton} ,
\end{align}
where $\vec{g}\of{k}_m$ and $\vec{H}\of{k}_m$ are the gradient and Hessian of the objective function in \eqref{mlr} at $\hvec{z}\of{k}_m$,
and $\alpha\of{k}\in (0,1]$ is a stepsize.
From \eqref{softmax}, it can be seen that $\frac{\partial}{\partial z_i}\log \pyz(y|\vec{z}) = \delta_{y-i} - \pyz(i|\vec{z})$, and so
\begin{align}
\vec{g}\of{k}_m
&= 
\vec{u}(\hvec{z}\of{k}_m) 
- \vec{e}_{y_m} 
+ [\Qp_m]^{-1}(\hvec{z}\of{k}_m-\hvec{p}_m) ,
\label{eq:gradient}
\end{align}
where 
$\vec{e}_y$ denotes the $y$th column of $\vec{I}_D$ 
and 
$\vec{u}(\vec{z})\in\Real^{D\times 1}$ is defined elementwise as 
\begin{align}
[\vec{u}(\vec{z})]_i \defn \pyz(i|\vec{z}) .
\label{eq:u}
\end{align}
Similarly, it is known \cite{Bohning:AISM:92} that the Hessian takes the form
\begin{align}
\vec{H}\of{k}_m
&= 
\vec{u}(\hvec{z}_m)\vec{u}(\hvec{z}_m)\tran - \Diag\{\vec{u}(\hvec{z}_m)\} - [\Qp_m]^{-1} ,
\label{eq:Hessian}
\end{align}
which also provides the answer to \lineref{msaQz} of \algref{hygamp}.
Note that each iteration $k$ of \eqref{newton} requires a $D\times D$ matrix inverse for each $m$.

It is possible to circumvent the matrix inversion in \eqref{newton} via componentwise update, i.e.,
\begin{align}
\hat{z}_{md}\of{k+1} 
&= \hat{z}_{md}\of{k} - \alpha\of{k} g\of{k}_{md} / H\of{k}_{md} 
\label{eq:cwnewton} ,
\end{align}
where $g\of{k}_{md}$ and $H\of{k}_{md}$ are the first and second derivatives of the objective function in \eqref{mlr} with respect to $z_{d}$ at $\vec{z}=\hvec{z}\of{k}_m$.
From \eqref{gradient}-\eqref{Hessian}, it follows that 
\begin{align}
g\of{k}_{md}
&= 
\pyz(d|\hvec{z}\of{k}_m) 
- \delta_{y_m-d} 
+  \big[ [\Qp_m]^{-1} \big]_{:,d}\tran (\hvec{z}\of{k}_m-\hvec{p}_m) 
\label{eq:cwgradient}\\
H\of{k}_{md}
&= \pyz(d|\hvec{z}\of{k}_m)^2 - \pyz(d|\hvec{z}\of{k}_m) - \big[ [\Qp_m]^{-1} \big]_{dd}  
\label{eq:cwHessian} .
\end{align}

\subsection{HyGAMP summary} \label{sec:hygamp_summary}

In summary, the SPA and MSA variants of the HyGAMP algorithm provide tractable methods of approximating the posterior test-label probabilities $p_{\ry_t|\rvecy_{1:M}}\big(y_t\biggiv\vec{y}_{1:M};\vec{A}\big)$ and computing the MAP weight matrix $\hvec{X}=\arg\max_{\vec{X}} p_{\rvecy_{1:M},\rvecX}(\vec{y}_{1:M},\vec{X};\vec{A})$, respectively, under a separable likelihood \eqref{py|X} and a separable prior \eqref{pX}. 
In particular, HyGAMP attacks the high-dimensional inference problems of interest using a sequence of $M+N$ low-dimensional (in particular, $D$-dimensional) inference problems and $D\times D$ matrix inversions, as detailed in \algref{hygamp}.

As detailed in the previous subsections, however, these $D$-dimensional inference problems are non-trivial in the sparse MLR case, making HyGAMP computationally costly.
\textb{We refer the reader to \tabref{hy_comp} for a summary of the $D$-dimensional inference problems encountered in running SPA-HyGAMP or MSA-HyGAMP, as well as their associated computational costs.}
Thus, in the sequel, we propose a computationally efficient simplification of HyGAMP that, as we will see in \secref{sims}, compares favorably with existing state-of-the-art methods.

\begin{table}[t]
\color{\blue}
\footnotesize
\centering
\begin{tabular}{|c|c|c|c|}
\hline 
Algorithm  & Quantity & Method  & Complexity \\  \hline 
\multirow{4}{*}{\begin{tabular}{c}SPA-\\HyGAMP\end{tabular}} 
                   & $\hvec{x}$  &  CF  &  $O(D^3)$ \\ 
		   & $\Qx$  	 &  CF  &  $O(D^3)$ \\ 
		   & $\hvec{z}$  &  NI  &  $O(D^K)$ \\ 
		   & $\Qz$  	 &  NI  &  $O(D^K)$ \\ \hline
\multirow{4}{*}{\begin{tabular}{c}MSA-\\HyGAMP\end{tabular}}  
                   & $\hvec{x}$  &  MM  &  $O(KD^3)$ \\ 
   		   & $\Qx$  	 &  CF  &  $O(D^3)$ \\ 
   		   & $\hvec{z}$  &  CWN &  $O(KD^2 \!+\! D^3)$ \\ 
   		   & $\Qz$  	 &  CF  &  $O(D^3)$ \\  \hline
\end{tabular}
\captionsetup{font=\mycaptionfont}
\caption{\textb{A summary of the $D$-dimensional inference sub-problems encountered when running SPA-HyGAMP or MSA-HyGAMP, as well as their associated computational costs.  `CF' = `closed form', `NI' = `numerical integration', `MM' = `minorization-maximization', and `CWN' = `component-wise Newton's method'. For the NI method, $K$ denotes the number of samples per dimension, and for the MM and CWN methods $K$ denotes the number of iterations.}}
\label{tab:hy_comp}
\vspace{-2mm}
\end{table}

\section{SHyGAMP for Multiclass Classification}		\label{sec:shygamp}

As described in \secref{hygamp}, a direct application of HyGAMP to sparse MLR is computationally costly.
Thus, in this section, we propose a \emph{simplified HyGAMP} (SHyGAMP) algorithm for sparse MLR, whose complexity is greatly reduced.
The simplification itself is rather straightforward: we constrain the covariance matrices $\Qr_n$, $\Qx_n$, $\Qp_m$, and $\Qz_m$ to be diagonal.
In other words, 
\begin{align}
\Qr_n = \Diag\big\{ \qr_{n1},\dots, \qr_{nD}\big\} ,
\end{align}
and similar for $\Qx_n$, $\Qp_m$, and $\Qz_m$.
As a consequence, the $D\times D$ matrix inversions in lines~\ref{line:Qs} and \ref{line:Qr} of \algref{hygamp} each reduce to $D$ scalar inversions.
More importantly, the $D$-dimensional inference problems in lines~\ref{line:msaxhat}-\ref{line:spaQx} and \ref{line:msazhat}-\ref{line:spaQz} can be tackled using much simpler methods than those described in \secref{hygamp}, as we detail below.

\subsection{Scalar Variance Approximation}    \label{sec:uni_var}

We further approximate the SHyGAMP algorithm using the \emph{scalar variance} GAMP approximation from \cite{Rangan:ISIT:12},
which reduces the memory and complexity of the algorithm.
The scalar variance approximation first approximates the variances $\{\qx_{nd}\}$ by a value invariant to both $n$ and $d$, i.e., 
\begin{equation}
\qx \defn \frac{1}{ND} \sum_{n=1}^N \sum_{d=1}^D \qx_{nd}.
\end{equation}
Then, in line~\ref{line:Qp} in \algref{hygamp}, we use the approximation 
\begin{align}
\qp_{md} &\approx \sum_{n=1}^N A^2_{mn} \qx 
\stackrel{(a)}{\approx} \frac{\norm{\vec{A}}^2_{F}}{M} \qx \label{eq:qp_approx}
\defn \qp.
\end{align}
The approximation (a), after precomputing $\norm{\vec{A}}^2_{F}$, reduces the complexity of line~\ref{line:Qp} from $O(ND)$ to $O(1)$.
We next define 
\begin{equation}
\qs 
\defn \frac{1}{MD} \sum_{m=1}^M \sum_{d=1}^D \qs_{md} 
\label{eq:qs_approx}
\end{equation}
and in line~\ref{line:Qr} we use the approximation 
\begin{align}
\qr_{nd}  
&\approx \left( \sum_{m=1}^M A^2_{mn} \qs \right)^{-1} 
\approx \frac{N}{\qs \norm{\vec{A}}_F^2} 
\defn \qr.
\end{align}
The complexity of line~\ref{line:Qr} then simplifies from $O(MD)$ to $O(1)$.
For clarity, we note that after applying the scalar variance approximation, we have $\Qx_n = \qx \vec{I}_D \, \forall \, n$, and similar for $\Qr_n$, $\Qp_m$ and $\Qz_m$.

\subsection{Sum-product SHyGAMP: Inference of \texorpdfstring{$\vec{x}_n$}{x}} \label{sec:spa_shygamp_input}

With diagonal $\Qr_n$ and $\Qx_n$, the implementation of lines~\ref{line:spaxhat}-\ref{line:spaQx} is greatly simplified by choosing a sparsifying prior $p_{\rvecx}$ with the separable form
$p_{\rvecx}(\vec{x}_n)=\prod_{d=1}^D \pxscalar(x_{nd})$.
A common example is the Bernoulli-Gaussian (BG) prior
\begin{align}
\pxscalar(x_{nd})
&= (1-\beta_d)\delta(x_{nd}) + \beta_d\mc{N}(x_{nd};m_d,v_d\vec{I}).
\label{eq:BG}
\end{align}
For any separable $p_{\rvecx}$, lines~\ref{line:spaxhat}-\ref{line:spaQx} reduce to computing the mean and variance of the distribution
\begin{align} 
\pxrscalar(x_{nd}|\hat{r}_{nd};\qr_{nd}) 
&= \textstyle
\frac{\pxscalar(x_{nd})\normal(x_{nd};\hat{r}_{nd},\qr_{nd})}
        {\int \pxscalar(x'_{nd})\normal(x'_{nd};\hat{r}_{nd},\qr_{nd}) \deriv x'_{nd}} 
\label{eq:pxrscalar}.
\end{align} 
for all $n=1\dots N$ and $d=1\dots D$, as in the simpler GAMP algorithm \cite{Rangan:ISIT:11}.
With the BG prior \eqref{BG}, these quantities can be computed in closed form (see, e.g., \cite{Schniter:CISS:10}).

\subsection{Sum-product SHyGAMP: Inference of \texorpdfstring{$\vec{z}_m$}{z}} \label{sec:spa_shygamp_output}

With diagonal $\Qp_m$ and $\Qz_m$, the implementation of lines~\ref{line:spazhat}-\ref{line:spaQz} can also be greatly simplified.
Essentially, the problem becomes that of computing the scalar means and variances
\begin{align} 
\hat{z}_{md} 
&= C_m^{-1} \!\!\int_{\Real^D} \! z_{d} \, p_{\ry|\rvecz}(y_m|\vec{z}) \prod_{k=1}^D\mc{N}(z_k;\hat{p}_{mk},\qp_{mk}) \deriv \vec{z} 
\label{eq:zhat} \\
\qz_{md} 
&=  C_m^{-1} \!\!\int_{\Real^D} \! z_{d}^2 \, p_{\ry|\rvecz}(y_m|\vec{z}) \prod_{k=1}^D\mc{N}(z_{k};\hat{p}_{mk},\qp_{mk}) \deriv \vec{z} - \hat{z}_{md}^2
\label{eq:zvar} 
\end{align} 
for $m=1\dots M$ and $d=1\dots D$. 
Here, $p_{\ry|\rvecz}$ has the MLR form in \eqref{softmax} and $C_m$ is a normalizing constant defined as
\begin{align} 
C_m
&\defn \int_{\Real^D} p_{\ry|\rvecz}(y_m|\vec{z}) \,\prod_{k=1}^D\mc{N}(z_{k};\hat{p}_{mk},\qp_{mk}) \deriv \vec{z}. 
\label{eq:zconst}
\end{align} 
Note that the likelihood $p_{\ry|\rvecz}$ is not separable and so inference does not decouple across $d$, as it did in \eqref{pxrscalar}.
We now describe several approaches to computing \eqref{zhat}-\eqref{zvar}.

\subsubsection{Numerical integration} \label{sec:ni_spa_out}

A straightforward approach to (approximately) computing \eqref{zhat}-\eqref{zconst} is through numerical integration (NI).
For this, we propose to use a hyper-rectangular grid of $\vec{z}$ values where, for $z_d$, the interval $\left[\hat{p}_{md}-\alpha \sqrt{\qp_{md}},~ \hat{p}_{md} + \alpha \sqrt{\qp_{md}}\right]$ is sampled at $K$ equi-spaced points.
Because a $D$-dimensional numerical integral must be computed for each index $m$ and $d$, the complexity of this approach grows as $O(MDK^D)$, making it impractical unless $D$, the number of classes, is very small.

\subsubsection{Importance sampling} \label{sec:is_spa_out}

An alternative approximation of \eqref{zhat}-\eqref{zconst} can be obtained through importance sampling (IS)\cite[\S 11.1.4]{Bishop:Book:07}.
Here, we draw $K$ independent samples $\{\tvec{z}_m[k]\}_{k=1}^K$ from $\normal(\hvec{p}_m, \Qp_m)$ and compute
\begin{align}
C_m
&\approx \sum_{k=1}^K  \pyz(y_m | \tvec{z}_m[k]) \\
\hat{z}_{md} 
&\approx C_m^{-1} \sum_{k=1}^K \tilde{z}_{md}[k] \pyz(y_m | \tvec{z}_m[k]) \\
\qz_{md}
&\approx C_m^{-1} \sum_{k=1}^K \tilde{z}_{md}^2[k]  \pyz(y_m | \tvec{z}_m[k]) - \hat{z}_{md}^2
\end{align}
for all $m$ and $d$.
The complexity of this approach grows as $O(MDK)$.

\subsubsection{Taylor-series approximation} \label{sec:taylor_spa_out}

Another approach is to approximate the likelihood $\pyz$ using a second-order Taylor series (TS) about $\hvec{p}_m$, i.e., $\pyz(y_m | \vec{z}) \approx f_m(\vec{z};\hvec{p}_m)$ with
\begin{align}
f_m(\vec{z};\hvec{p}_m)
&\defn \pyz(y_m | \hvec{p}_m) +  \vec{g}_m(\hvec{p}_m)\tran(\vec{z} - \hvec{p}_m) 
\nonumber \\ & \quad
+ \frac{1}{2}(\vec{z} - \hvec{p}_m)\tran  \vec{H}_m(\hvec{p}_m) (\vec{z} - \hvec{p}_m) 
\label{eq:like_taygen}
\end{align}
for gradient $\vec{g}_m(\hvec{p}) \defn \frac{\partial}{\partial \vec{z}} \pyz(y_m | \vec{z})\barst{\vec{z}=\hvec{p}}$ and Hessian $\vec{H}_m(\hvec{p}) \defn \frac{\partial^2}{\partial \vec{z}^2} \pyz(y_m | \vec{z})\barst{\vec{z}=\hvec{p}}$.
In this case, it can be shown \cite{Byrne:Thesis:15} that 
\begin{align}
C_m 
&\approx f_m(\hvec{p}_m) + \frac{1}{2} \sum_{k=1}^D H_{mk}(\hvec{p}_m)\qp_{mk} 
\label{eq:taylor_c} \\
\hat{z}_{md} 
&\approx \hat{C}_m^{-1} \Bigg( f_m(\hvec{p}_m) \, \hat{p}_{md}^{\phantom{1}} +  g_{md}(\hvec{p}_m) \qp_{md} 
\nonumber \\ &\quad
+ \frac{1}{2} \sum_{k=1}^D \hat{p}_{mk}\qp_{mk} H_{mk}(\hvec{p}_m) \Bigg) 
\label{eq:taylor_mean} \\
\qz_{md} 
&\approx C_m^{-1} \Bigg( f_m(\hvec{p}_m) \, ( \hat{p}_{md}^2 + \qp_{md}) +  2 g_{md}(\hvec{p}_m) \hat{p}_{md} \qp_{md}  
\nonumber \\ & \quad
+ \frac{1}{2} \qp_{md} \big(\hat{p}_{md}^2 + 3 \qp_{md} \big) H_{md}(\hvec{p}_m) 
\nonumber \\& \quad
+ \frac{1}{2}  \big(\hat{p}_{md}^2 + \qp_{md}\big) H_{md}(\hvec{p}_m) 
\sum_{k\neq d} \qp_{mk} \Bigg)  -  \hat{z}_{md}^2  
\label{eq:taylor_var} ,
\end{align}
where $H_{md}(\hvec{p})\defn [\vec{H}_m(\hvec{p})]_{dd}$.
The complexity of this approach grows as $O(MD)$.

\subsubsection{Gaussian mixture approximation} \label{sec:gm_spa_out}

It is known that the logistic cdf $1/(1+\exp(-x))$ is well approximated by a mixture of a few Gaussian cdfs, which leads to an efficient method of approximating \eqref{zhat}-\eqref{zvar} in the case of \emph{binary} logistic regression (i.e., $D=2$) \cite{Stefanski:SPL:91}.
We now develop an extension of this method for the MLR case (i.e., $D\geq 2$).

To facilitate the Gaussian mixture (GM) approximation, we work with the difference variables 
\begin{align}
\gamma_d\of{y} 
\defn \begin{cases} z_y - z_d & d\neq y\\ z_y & d=y \end{cases}.
\end{align}
Their utility can be seen from the fact that (recalling \eqref{softmax})
\begin{align}
p_{\ry|\rvecz}(y|\vec{z}) 
&= \frac{1}{1 + \sum_{d\neq y} \exp (z_d-z_y)} \\
&= \frac{1}{1 + \sum_{d\neq y} \exp(-\gamma_d\of{y})} 
\defn l\of{y}(\vec{\gamma}\of{y})
\label{eq:mnl_e} ,
\end{align}
which is smooth, positive, and bounded by $1$, and strictly increasing in $\gamma_d\of{y}$. 
Thus,\footnote{Note that, since the role of $y$ in $\hat{l}\of{y}(\vec{\gamma})$ is merely to ignore the $y$th component of the input $\vec{\gamma}$, we could have instead written
$\hat{l}\of{y}(\vec{\gamma}) = \hat{l}(\vec{J}_y\vec{\gamma})$
for $y$-invariant $\hat{l}(\cdot)$ and $\vec{J}_y$ constructed by removing the $y$th row from the identity matrix.}
for appropriately chosen $\{\alpha_l, \mu_{kl}, \sigma_{kl}\}$,
\begin{align}
l\of{y}(\vec{\gamma}) 
&\approx \sum_{l=1}^L \alpha_l \prod_{k\neq y} 
\Phi\bigg(\frac{\gamma_k - \mu_{kl}}{\sigma_{kl}}\bigg)
\defn \hat{l}\of{y}(\vec{\gamma}),
\label{eq:gm}
\end{align}
where $\Phi(x)$ is the standard normal cdf, $\sigma_{kl}>0$, $\alpha_l \geq 0$, and $\sum_l \alpha_l = 1$.
In practice, the GM parameters $\{\alpha_l,\mu_{kl},\sigma_{kl}\}$ could be designed off-line to minimize, e.g., the total variation distance $\sup_{\vec{\gamma}\in\Real^D}|l\of{y}(\vec{\gamma})-\hat{l}\of{y}(\vec{\gamma})|$.

Recall from \eqref{zhat}-\eqref{zconst} that our objective is to compute quantities of the form
\begin{align}
\int_{\Real^D} (\vec{e}_d\tran \vec{z})^i \, \pyz(y|\vec{z}) \mc{N}(\vec{z};\hvec{p},\Qp) \deriv\vec{z}
\defn 
S_{di}\of{y} ,
\label{eq:Sdiy}
\end{align}
where $i\in\{0,1,2\}$, $\Qp$ is diagonal, and $\vec{e}_d$ is the $d$th column of $\vec{I}_D$.
To exploit \eqref{gm}, we change the integration variable to
\begin{align}
\vec{\gamma}\of{y}
&= \vec{T}_y \vec{z}
\label{eq:Ty} 
\end{align}
with
\begin{align}
\vec{T}_y 
&= \mat{-\vec{I}_{y-1} & \vec{1}_{(y-1)\times 1} & \vec{0}_{(y-1)\times(D-y)}\\
        \vec{0}_{1\times (y-1)} & 1 & \vec{0}_{1\times (D-y)} \\
        \vec{0}_{(D-y)\times(y-1)} & \vec{1}_{(D-y)\times 1} & -\vec{I}_{D-y}\\
        } 
\end{align}
to get
(since $\det(\vec{T}_y)=1$)
\begin{align}
S_{di}\of{y}
&= \int_{\Real^D} \big(\vec{e}_d\tran \vec{T}_y^{-1} \vec{\gamma}\big)^i \, 
l\of{y}(\vec{\gamma}) 
\mc{N}\big(\vec{\gamma};\vec{T}_y\hvec{p},\vec{T}_y\Qp\vec{T}_y\tran\big) 
\deriv\vec{\gamma} .
\label{eq:Sdiy2}
\end{align}
Then, applying the approximation \eqref{gm} and 
\begin{align}
\mc{N}\big(\vec{\gamma};\vec{T}_y\hvec{p},\vec{T}_y\Qp\vec{T}_y\tran\big)
&= \mc{N}\big(\gamma_y; \hat{p}_y,\qp_y\big) 
\nonumber \\ & \quad \times
\prod_{k\neq y} \mc{N}\big(\gamma_k;\gamma_y-\hat{p}_k,\qp_k\big) 
\end{align}
to \eqref{Sdiy2}, we find that 
\begin{align}
S_{di}\of{y}
&\approx \sum_{l=1}^L \alpha_l 
\int_\Real
\mc{N}\big(\gamma_y; \hat{p}_y,\qp_y\big) 
\bigg[ 
\int_{\Real^{D-1}}
\big(\vec{e}_d\tran \vec{T}_y^{-1} \vec{\gamma}\big)^i \,
\nonumber \\ & \quad \times
\prod_{k\neq y} 
\mc{N}\big(\gamma_k;\gamma_y-\hat{p}_k,\qp_k\big) 
\Phi\bigg(\frac{\gamma_k - \mu_{kl}}{\sigma_{kl}}\bigg) 
\deriv\gamma_k 
\bigg]
\deriv\gamma_y .
\label{eq:Sdiy3}
\end{align}
Noting that $\vec{T}_y^{-1}=\vec{T}_y$, we have
\begin{align}
\vec{e}_d\tran \vec{T}_y^{-1} \vec{\gamma}
&= \begin{cases}
\gamma_y-\gamma_d & d\neq y \\
\gamma_y & d=y \\
\end{cases} .
\end{align}
Thus, for a fixed value of $\gamma_y=c$, the inner integral in \eqref{Sdiy3} can be expressed as a product of linear combinations of terms 
\begin{align}
\int_\Real
\gamma^i \,
\mc{N}\big(\gamma;c-\hat{p},q\big) 
\Phi\bigg(\frac{\gamma - \mu}{\sigma}\bigg) 
\deriv\gamma 
\defn T_i
\end{align}
with $i\in\{0,1,2\}$, which can be computed in closed form.
In particular, defining 
$x\defn \frac{c-\hat{p}-\mu}{\sqrt{\sigma^2 + q}}$, we have
\begin{align}
T_0
&= \Phi(x) \\
T_1
&= (c-\hat{p}) \Phi(x) + \frac{q \phi(x)}{\sqrt{\sigma^2 + q}} \\
T_2
&= \frac{(T_1)^2}{\Phi(x)} + q\Phi(x) - \frac{q^2 \phi(x)}{\sigma^2 + q}\Big(x+\frac{\phi(x)}{\Phi(x)}\Big) ,
\end{align}
which can be obtained using the results in \cite[\S 3.9]{Rasmussen:Book:06}.
The outer integral in \eqref{Sdiy3} can then be approximated via numerical integration.

If a grid of $K$ values is used for numerical integration over $\gamma_y$ in \eqref{Sdiy3}, then the overall complexity of the method grows as $O(MDLK)$.
Our experiments indicate that relatively small values (e.g., $L=2$ and $K=7$) suffice.

\subsubsection{Performance comparison} \label{sec:compare_spa_out}

\newcommand{\true}{_{\textsf{true}}}
\newcommand{\MSE}{\text{MSE}}

Above we described four methods of approximating lines~\ref{line:spazhat}-\ref{line:spaQz} in \algref{hygamp} under diagonal $\Qp$ and $\Qz$. 
We now compare the accuracy and complexity of these methods.
In particular, we measured the accuracy of the conditional mean (i.e., \lineref{spazhat}) approximation as follows (for a given $\hvec{p}$ and $\Qp$):
\begin{enumerate}
\item generate i.i.d.\ samples $\vec{z}\true[t] \sim \mc{N}(\vec{z};\hvec{p}, \Qp)$ and $y\true[t] \sim \pyz(y \giv \vec{z}\true[t])$ for $t=1\dots T$,
\item compute the approximation $\hvec{z}[t] \approx \E\{\rvecz \giv \ry = y\true[t],\rvecp = \hvec{p}; \Qp\}$ using each method described in Sections~\ref{sec:ni_spa_out}--\ref{sec:gm_spa_out}, 
\item compute average $\MSE \defn \frac{1}{T} \sum_{t=1}^T \big\|\vec{z}\true[t] - \hvec{z}[t]\big\|_2^2$ for each method,
\end{enumerate}
and we measured the combined runtime of lines~\ref{line:spazhat}-\ref{line:spaQz} for each method.
Unless otherwise noted, we used $D=4$ classes, $\hvec{p}=\vec{e}_1$, $\Qp=\qp\vec{I}_D$, and $\qp=1$ in our experiments.
For numerical integration (NI), we used a grid of size $K = 7$ and radius of $\alpha = 4$ standard deviations; 
for importance sampling (IS), we used $K = 1500$ samples; and
for the Gaussian-mixture (GM) method, we used $L = 2$ mixture components and a grid size of $K = 7$.
Empirically, we found that smaller grids or fewer samples compromised accuracy, whereas larger grids or more samples compromised runtime.

\begin{figure}[t]
\centering
\psfrag{pvar}[ct][c][0.8]{\textsf{Variance ${\qp}$}}
\psfrag{mse}[cb][c][0.8]{\textsf{MSE$/{\qp}$}}
\psfrag{5e6 trials}{}
\psfrag{Method1}[l][][.75]{Numerical Int}
\psfrag{Method2}[l][][.75]{Importance Sampling}
\psfrag{Method3}[l][][.75]{Taylor Series}
\psfrag{Method4}[l][][.75]{Gaussian Post.}
\psfrag{Method 5}[l][][.75]{Gaussian Mix.}
\psfrag{trivial}[l][][.75]{Trivial}
\includegraphics[width=\figsize]{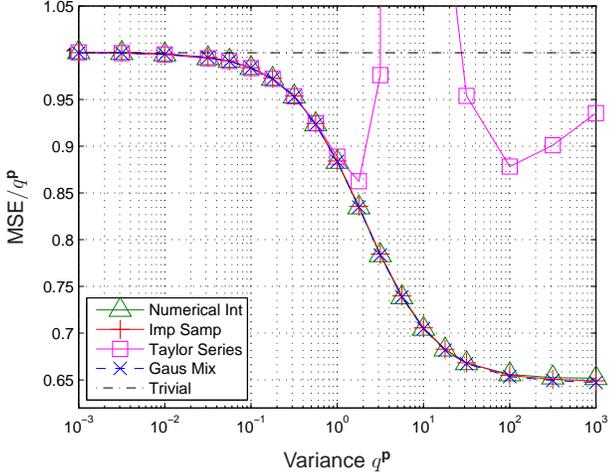}
\captionsetup{font=\mycaptionfont}
\caption{$\MSE/\qp$ versus variance $\qp$ for various methods to compute \lineref{spazhat} in \algref{hygamp}.  Each point represents the average of $5\times 10^6$ independent trials. 
}
\label{fig:msevpvar1}
\end{figure}

\Figref{msevpvar1} plots the normalized MSE versus variance $\qp$ for the four methods under test, in addition to the trivial method $\hvec{z}[t] = \hvec{p}$. 
The figure shows that the NI, IS, and GM methods performed similarly across the full range of $\qp$ and always outperform the trivial method. 
The Taylor-series method, however, breaks down when $\qp>1$.
A close examination of the figure reveals that GM gave the best accuracy, IS the second best accuracy, and NI the third best accuracy.

\begin{figure}[t]
\centering
\psfrag{runtime}[cb][c][0.8]{\textsf{Runtime [sec]}}
\psfrag{D}[ct][c][0.8]{\textsf{Number of Classes $D$}}
\psfrag{trials = 1000000}[l][][.75]{ }
\includegraphics[width=\figsize]{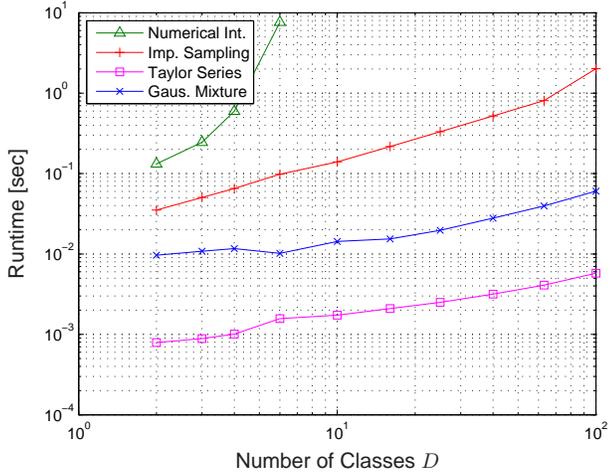}
\captionsetup{font=\mycaptionfont}
\caption{Cumulative runtime (over $M=500$ samples) versus number-of-classes $D$ for various methods to compute lines~\ref{line:spazhat}-\ref{line:spaQz} in \algref{hygamp}.  Each point represents the average of $2000$ independent trials. 
}
\label{fig:mmse_estim_runtime}
\end{figure}

\Figref{mmse_estim_runtime} shows the cumulative runtime (over $M=500$ training samples) of the methods from Sections~\ref{sec:ni_spa_out}--\ref{sec:gm_spa_out} versus the number of classes, $D$.
Although the Taylor-series method was the fastest, we saw in \figref{msevpvar1} that it is accurate only at small variances $\qp$.
\Figref{mmse_estim_runtime} then shows GM was about an order-of-magnitude faster than IS, which was several orders-of-magnitude faster than NI.

Together, Figures~\ref{fig:msevpvar1}-\ref{fig:mmse_estim_runtime}, show that our proposed GM method dominated the IS and NI methods in both accuracy and runtime.
Thus, for the remainder of the paper, we implement sum-product SHyGAMP using the GM method from \secref{gm_spa_out}.

\subsection{Min-sum SHyGAMP: Inference of \texorpdfstring{$\vec{x}_n$}{x}}  \label{sec:msa_shygamp_input}

With diagonal $\Qr_n$ and $\Qx_n$, the implementation of lines~\ref{line:msaxhat}-\ref{line:msaQx} in \algref{hygamp} can be significantly simplified.
Recall that, when the prior $\px$ is chosen as i.i.d.\ Laplace \eqref{laplace}, \lineref{msaxhat} manifests as \eqref{lasso}, which is in general a non-trivial optimization problem.
But with diagonal $\Qr_n$, \eqref{lasso} decouples into $D$ instances of the scalar optimization
\begin{align}
x_{nd}
&= \arg\max_{x}
-\frac{1}{2} \frac{(x-\hat{r}_{nd})^2}{\qr_{nd}} - \lambda |x| 
\label{eq:scalar_lasso} ,
\end{align}
which is known to have the closed-form ``soft thresholding'' solution
\begin{align}
\hat{x}_{nd} 
= \sgn(\hat{r}_{nd})  \max \{ 0, |\hat{r}_{nd}| - \lambda \qr_{nd} \}. 
\label{eq:softthresh}
\end{align}
Above, $\sgn(r)=1$ when $r\geq 0$ and $\sgn(r)=-1$ when $r<0$.

Meanwhile, \lineref{msaQx} reduces to
\begin{align}
\qx_{nd}
&= \left[ \left. \frac{\partial^2}{\partial x^2} \left( \frac{1}{2} \frac{(x-\hat{r}_{nd})^2}{\qr_{nd}} + \lambda |x| \right)\right|_{x=\hat{x}_{nd}} \right]^{-1} ,
\end{align}
which equals $\qr_{nd}$ when $\hat{x}_{nd}\neq 0$ and is otherwise undefined. 
When $\hat{x}_{nd}=0$, we set $\qx_{nd}=0$.

\subsection{Min-sum SHyGAMP: Inference of \texorpdfstring{$\vec{z}_m$}{z}}  \label{sec:msa_shygamp_output}

With diagonal $\Qp_m$ and $\Qz_m$, the implementation of lines~\ref{line:msazhat}-\ref{line:msaQz} in \algref{hygamp} also simplifies.
Recall that, when the likelihood $\pyz$ takes the MLR form in \eqref{softmax}, \lineref{msazhat} manifests as \eqref{mlr}, which can be solved using a component-wise Newton's method as in \eqref{cwnewton}-\eqref{cwHessian} for any $\Qp_m$ and $\Qz_m$.
When $\Qp_m$ is diagonal, the first and second derivatives \eqref{cwgradient}-\eqref{cwHessian} reduce to 
\begin{align}
g\of{k}_{md}
&= 
\pyz(d|\hvec{z}\of{k}_m) 
- \delta_{y_m-d} 
+  (\hat{z}\of{k}_{md}-\hat{p}_{md})/\qp_{md} .
\label{eq:cwgradient2} \\
H\of{k}_{md}
&= \pyz(d|\hvec{z}\of{k}_m)^2 - \pyz(d|\hvec{z}\of{k}_m) -  1/\qp_{md}  
\label{eq:cwHessian2} ,
\end{align}
which leads to a reduction in complexity.

Furthermore, \lineref{msaQz} simplifies, since with diagonal $\Qz_m$ it suffices to compute only the diagonal components of $\vec{H}\of{k}_m$ in \eqref{Hessian}.
In particular, \textb{when $\Qp_m$ is diagonal}, the result becomes 
\begin{align}
\qz_{md}
&= \frac{1}{1/\qp_{md} + \pyz(d|\hvec{z}_m) - \pyz(d|\hvec{z}_m)^2 }.
\end{align}

\subsection{SHyGAMP summary}

\color{\blue}
In summary, by approximating the covariance matrices as diagonal, the SPA-SHyGAMP and MSA-SHyGAMP algorithms improve computationally upon their HyGAMP counterparts.
A summary of the $D$-dimensional inference problems encountered when running SPA-SHyGAMP or MSA-SHyGAMP, as well as their associated computational costs, is given in \tabref{shy_comp}.
A high-level comparison between HyGAMP and SHyGAMP is given in \tabref{comp1}.

\begin{table}[t]
\footnotesize
\color{\blue}
\centering
\begin{tabular}{|c|c|c|c|}
\hline 
Algorithm  & Quantity & Method  & Complexity \\  \hline 
\multirow{4}{*}{\begin{tabular}{c}SPA-\\SHyGAMP\end{tabular}} 
                    & $\hvec{x}$  &  CF  &  $O(D)$  \\ 
		    & $\Qx$  	  &  CF  &  $O(D)$  \\ 
		    & $\hvec{z}$  &  GM  &  $O(LKD)$\\ 
		    & $\Qz$  	  &  GM  &  $O(LKD)$\\ \hline
\multirow{4}{*}{\begin{tabular}{c}MSA-\\SHyGAMP\end{tabular}} 
                    & $\hvec{x}$  &  ST  &  $O(D)$  \\ 
   		    & $\Qx$  	  &  CF  &  $O(D)$  \\ 
   		    & $\hvec{z}$  &  CWN &  $O(KD)$ \\ 
   		    & $\Qz$  	  &  CF  &  $O(D^3)$\\  \hline
\end{tabular}
\captionsetup{font=\mycaptionfont}
\caption{\textb{A summary of the $D$-dimensional inference sub-problems encountered when running SPA-SHyGAMP or MSA-SHyGAMP, as well as their associated computational costs.  `CF' = `closed form', `GM' = `Gaussian mixture', `ST' = `Soft-thresholding', and `CWN' = `component-wise Newton's method'. For the GM, $L$ denotes the number of mixture components and $K$ the number of samples in the 1D numerical integral, and for CWN $K$ denotes the number of iterations.}}
\label{tab:shy_comp}
\end{table}

\begin{table}[t]
\color{\blue}
\footnotesize
\centering
\begin{tabular}{|l|c|c|}
\hline
Algorithm & HyGAMP & SHyGAMP \\ \hline
Diagonal covariance matrices & & $\checkmark$ \\
Simplified $D$-dimensional inference & & $\checkmark$ \\
Scalar-variance approximation & & $\checkmark$ \\
Online parameter tuning & & $\checkmark$ \\
\hline
\end{tabular}

\captionsetup{font=\mycaptionfont}
\caption{\textb{High-level comparison of SHyGAMP and HyGAMP.}}
\label{tab:comp1}
\end{table}

\color{black}

\section{Online Parameter Tuning}		\label{sec:param_tune}

The weight vector priors in \eqref{laplace} and \eqref{BG} depend on modeling parameters that, in practice, must be tuned.
Although cross-validation (CV) is the customary approach to tuning \textb{such} model parameters, it 
can be very computationally costly, since each parameter must be tested over a grid of hypothesized values and over multiple data folds.  
For example, $K$-fold cross-validation tuning of $P$ parameters using $G$ hypothesized values of each parameter requires the training and evaluation of $KG^P$ classifiers. 

\subsection{Parameter selection for Sum-product SHyGAMP}  \label{sec:param_tune_EM}

For SPA-SHyGAMP, we propose to use the zero-mean Bernoulli-Gaussian prior in \eqref{BG}, which has parameters $\beta_d$, $m_d$, and $v_d$.
Instead of CV, we use the EM-GM-AMP framework described in \cite{Vila:TSP:13} to tune these parameters online.
See \cite{Byrne:Thesis:15} for details regarding the initialization of $\beta_d$, $m_d$, and $v_d$.

\subsection{Parameter selection for Min-sum SHyGAMP}   \label{sec:param_tune_SURE}

To use MSA-SHyGAMP with the Laplacian prior in \eqref{laplace}, we need to specify the scale parameter $\lambda$.
For this, we use a modification of the SURE-AMP framework from \cite{Mousavi:13}, which adjusts $\lambda$ to minimize the Stein's unbiased risk estimate (SURE) of the weight-vector MSE.

We describe our method by first reviewing SURE and SURE-AMP.
First, suppose that the goal is to estimate the value of $x$,      
which is a realization of the random variable $\textsf{x}$,
from the noisy observation $r$,       
which is a realization of 
\begin{equation}
\textsf{r}=\textsf{x}+\sqrt{\qr}\textsf{w},
\label{eq:sure_assum}
\end{equation}
with $\textsf{w}\sim\mc{N}(0,1)$ and $\qr>0$.
For this purpose, consider an estimate of the form
$\hat{x}=f(r, \qr ; \vec{\theta})$
where $\vec{\theta}$ contains tunable parameters.
For convenience, define the shifted estimation function
$g(r,\qr;\vec{\theta}) \defn f(r,\qr;\vec{\theta}) - r$ 
and its derivative $g'(r,\qr;\vec{\theta}) \defn \frac{\partial}{\partial r} g(r,\qr;\vec{\theta})$.
Then Stein \cite{Stein:AS:81} established the following result on the mean-squared error, or risk, of the estimate $\hat{x}$:
\begin{align}
\E\big\{[\hat{\textsf{x}}-\textsf{x}]^2\big\}
&= \qr + \E\big\{g^2(\textsf{r},\qr;\vec{\theta}) + 2\qr g'(\textsf{r},\qr;\vec{\theta})\big\} .
\label{eq:sure}
\end{align}
The implication of \eqref{sure} is that, given only the noisy observation $r$ and the noise variance $\qr$, one can compute an estimate
\begin{align}
\text{SURE}(r,\qr;\vec{\theta}) 
&\defn \qr + g^2\big(r,\qr;\vec{\theta}) + 2 \qr g'(r,\qr;\vec{\theta}) 
\label{eq:sure_defn}
\end{align}
of the $\text{MSE}(\vec{\theta})\defn \E\big\{[\hat{\textsf{x}}-\textsf{x}]^2\big\}$
that is unbiased, i.e.,
\begin{align}
\E\big\{ \text{SURE}(\textsf{r},\qr;\vec{\theta}) \big\}
&= \text{MSE}(\vec{\theta}).
\end{align}
These unbiased risk estimates can then be used as a surrogate for the true MSE when tuning $\vec{\theta}$.

In \cite{Mousavi:13}, it was noticed that the assumption \eqref{sure_assum} is satisfied by AMP's denoiser inputs $\{\hat{r}_n\}_{n=1}^N$, and thus \cite{Mousavi:13} proposed to tune the soft threshold $\lambda$ to minimize the SURE:
\begin{equation}
\hat{\lambda} 
= \argmin_{\lambda} \sum_{n=1}^N g^2\big(\hat{r}_{n},\qr ; \lambda) + 2 \qr g'(\hat{r}_{n}, \qr ; \lambda ).
\label{eq:sure_cost}
\end{equation}
Recalling the form of the estimator $f(\cdot)$ from \eqref{softthresh}, we have
\begin{align}
g^2 (\hat{r}_{n}, \qr ; \lambda) 
&= \begin{cases} \lambda^2 (\qr)^2 & \text{if~} |\hat{r}_{n}| > \lambda \qr \\ \hat{r}_{n}^2 & \text{ otherwise} \end{cases} 
\label{eq:sure_g2}\\
g'( \hat{r}_{n}, \qr ; \lambda ) 
&= \begin{cases} -1 & \text{if~} |\hat{r}_{n}| < \lambda \qr \\ 0 & \text{ otherwise} \end{cases}.
\label{eq:sure_g'}
\end{align}
However, solving \eqref{sure_cost} for $\lambda$ is non-trivial because the objective is non-smooth and has many local minima.
A stochastic gradient descent approach was proposed in \cite{Mousavi:13}, but its convergence speed is too slow to be practical.

Since \eqref{sure_assum} also matches the scalar-variance SHyGAMP model from \secref{uni_var}, we propose to use SURE to tune $\lambda$ for min-sum SHyGAMP.
But, instead of the empirical average in \eqref{sure_cost}, we propose to use a statistical average, i.e.,
\begin{equation}
\hat{\lambda} 
= \argmin_{\lambda} \underbrace{
\E\big\{ g^2\big(\textsf{r},\qr ; \lambda) + 2 \qr g'(\textsf{r}, \qr ; \lambda ) \big\}
}_{\displaystyle \defn J(\lambda)}
\label{eq:sure_cost2} ,
\end{equation}
by modeling the random variable $\textsf{r}$ as a Gaussian mixture (GM) whose parameters are fitted to $\{\hat{r}_{nd}\}$.
As a result, the objective in \eqref{sure_cost2} is smooth.
Moreover, by constraining the smallest mixture variance to be at least $\qr$, the objective becomes unimodal, in which case $\hat{\lambda}$ from \eqref{sure_cost2} is the unique root of $\frac{\deriv}{\deriv\lambda}J(\lambda)$.
To find this root, we use the bisection method.
In particular, due to \eqref{sure_g2}-\eqref{sure_g'},
the objective in \eqref{sure_cost2} becomes
\begin{align}
J(\lambda)
&=\int_{-\infty}^{-\lambda \qr} \pr \lambda^2 (\qr)^2 \deriv r + \int_{-\lambda \qr}^{\lambda \qr} \pr (r^2 - 2 \qr) \deriv r  
\nonumber \\ &\quad 
+ \int_{\lambda \qr}^{\infty} \pr \lambda^2 (\qr)^2 \deriv r ,
\label{eq:sure_cost_lambda}
\end{align}
from which it can be shown that \cite{Byrne:Thesis:15}
\begin{align}
\frac{\deriv}{\deriv \lambda} J(\lambda)
&= 2 \lambda (\qr)^2 \big[ 1 - \Pr\{-\lambda\qr < \textsf{r} < \lambda \qr\}\big]
\nonumber \\ &\quad 
- \big[p_{\textsf{r}}(\lambda \qr)+p_{\textsf{r}}(-\lambda \qr)\big] 2(\qr)^2 
\label{eq:gm_grad} .
\end{align}
For GM fitting, we use the standard EM approach \cite{Bishop:Book:07} and find that relatively few (e.g., $L=3$) mixture terms suffice.
Note that we re-tune $\lambda$ using the above technique at each iteration of \algref{hygamp}, immediately before line~\ref{line:msaxhat}.
Experimental verification of our method is provided in \secref{sims_sure}.

\section{Numerical Results}		\label{sec:sims}


In this section we describe the results of several experiments used to test SHyGAMP.
In these experiments, EM-tuned SPA-SHyGAMP and SURE-tuned MSA-SHyGAMP were compared to two state-of-the-art sparse MLR algorithms: SBMLR\cite{Cawley:NIPS:07} and GLMNET\cite{Friedman:JSS:10}.
We are particularly interested in SBMLR and GLMNET because \cite{Cawley:NIPS:07,Friedman:JSS:10} show that they have strong advantages over earlier algorithms, e.g., \cite{Tipping:JMLR:01,Krishnapuram:TPAMI:05,Genkin:Techno:07}.
As described in \secref{existing}, both SBMLR and GLMNET use $\ell_1$ regularization, but SBMLR tunes the regularization parameter $\lambda$ using evidence maximization while GLMNET tunes it using cross-validation \textb{(using the default value of $10$ folds unless otherwise noted)}. 
For SBMLR and GLMNET, we ran code written by the authors%
\footnote{SBMLR obtained from \url{http://theoval.cmp.uea.ac.uk/matlab/}}%
\footnote{GLMNET obtained from \url{http://www.stanford.edu/~hastie/glmnet_matlab/}} 
under default settings \textb{(unless otherwise noted)}.
For SHyGAMP, we used the damping modification described in \cite{Rangan:ISIT:14}.
We note that the runtimes reported for all algorithms include the total time spent to tune all parameters and train the final classifier.


Due to space limitations, we do not show the performance of the more complicated HyGAMP algorithm from \secref{hygamp}.
However, our experience suggests that HyGAMP generates weight matrices $\hvec{X}$ that are very similar to those generated by SHyGAMP, but with much longer runtimes, especially as $D$ grows.

\subsection{Synthetic data in the \texorpdfstring{$M\ll N$}{undersampled} regime} \label{sec:sims_synth}

We first describe the results of three experiments with synthetic data. 
For these experiments, the training data \textb{were} randomly generated and algorithm performance was averaged over several data realizations. 
In all cases, we started with balanced training labels $y_m \in \{1,\dots,D\}$ for $m=1,\dots,M$ (i.e., $M/D$ examples from each of $D$ classes).
Then, for each data realization, we generated $M$ i.i.d.\ training features $\vec{a}_m$ from the class-conditional generative distribution $\vec{a}_m \giv y_m \sim \mc{N}(\vec{\mu}_{y_m}, v \vec{I}_N)$.
In doing so, we chose the intra-class variance, $v$, to attain a desired Bayes error rate (BER) of $10\%$ (see \cite{Byrne:Thesis:15} for details), and we used randomly generated $K$-sparse orthonormal class means, $\vec{\mu}_{d}\in\Real^N$.
In particular, we generated $[\vec{\mu}_1,\dots,\vec{\mu}_D]$ by drawing a $K\times K$ matrix with i.i.d.\ $\mc{N}(0,1)$ entries, performing a singular value decomposition, and zero-padding the first $D$ left singular vectors to length $N$.
We note that our generation of $\vec{y},\vec{A},\vec{X}$ is matched \cite{Jordan:Tech:95} to the multinomial logistic model \eqref{py|X}-\eqref{softmax}.

Given a training data realization, each algorithm was invoked to yield a weight matrix $\hvec{X}=[\hvec{x}_1,\dots,\hvec{x}_D]$.
The corresponding \textb{expected} test-error rate was then analytically computed as 
\begin{eqnarray}
\Pr\{\text{err}\}
&=& 1-\frac{1}{D}\sum_{y=1}^D \Pr\{\text{cor}|y\} 
\label{eq:Perr}\\
\Pr\{\text{cor}|y\}
&=& \Pr \bigcap_{d\neq y} \left\{ (\hvec{x}_y-\hvec{x}_d)\tran\rveca < (\hvec{x}_y-\hvec{x}_d)\tran\vec{\mu}_y \right\} ,
\qquad
\label{eq:Perry}
\end{eqnarray}
where $\rveca\sim\mc{N}(\vec{0},v\vec{I}_N)$ and the multivariate normal cdf in \eqref{Perry} was computed using Matlab's \texttt{mvncdf}.

For all three synthetic-data experiments, we used $D=4$ classes and $K\ll M\ll N$. 
In the first experiment, we fixed $K$ and $N$ and we varied $M$; 
in the second experiment, we fixed $K$ and $M$ and we varied $K$; and 
in the third experiment, we fixed $K$ and $M$ and we varied $N$.
The specific values/ranges of $K,M,N$ used for each experiment are given in \tabref{syn_param}.
\begin{table}[t]
\centering
\scriptsize
\begin{tabular}{|c|c|c|c|c|}
\hline 
Experiment & $M$                 &  $N$                      & $K$           & $D$ \\  \hline 
1          & $\{100,\dots,5000\}$&  $10000$                  & $10$          & $4$ \\ 
2          & $300$               &  $30000$                  & $\{5,\dots,30\}$ & $4$ \\ 
3          & $200$               &  $\{10^3,\dots,10^{5.5}\}$& $10$          & $4$ \\ 
\textb{4}  & \textb{$300$}       &  \textb{$30000$}          & \textb{$25$}  & \textb{$4$} \\ \hline
\end{tabular}
\captionsetup{font=\mycaptionfont}
\caption{Configurations of the synthetic-data experiments.}
\label{tab:syn_param}
\end{table}

\begin{figure}
\captionsetup{font=\mycaptionfont}
\centering
\begin{minipage}{\figsize}
\psfrag{D = 4, N = 10000, K = 10, BER = 0.10,trials = 12}{}
\psfrag{error}[b][b][0.8]{\textsf{Test Error Rate}}
\psfrag{M}[t][t][0.8]{\textsf{Number of Training Samples $M$}}
\includegraphics[width=1\textwidth]{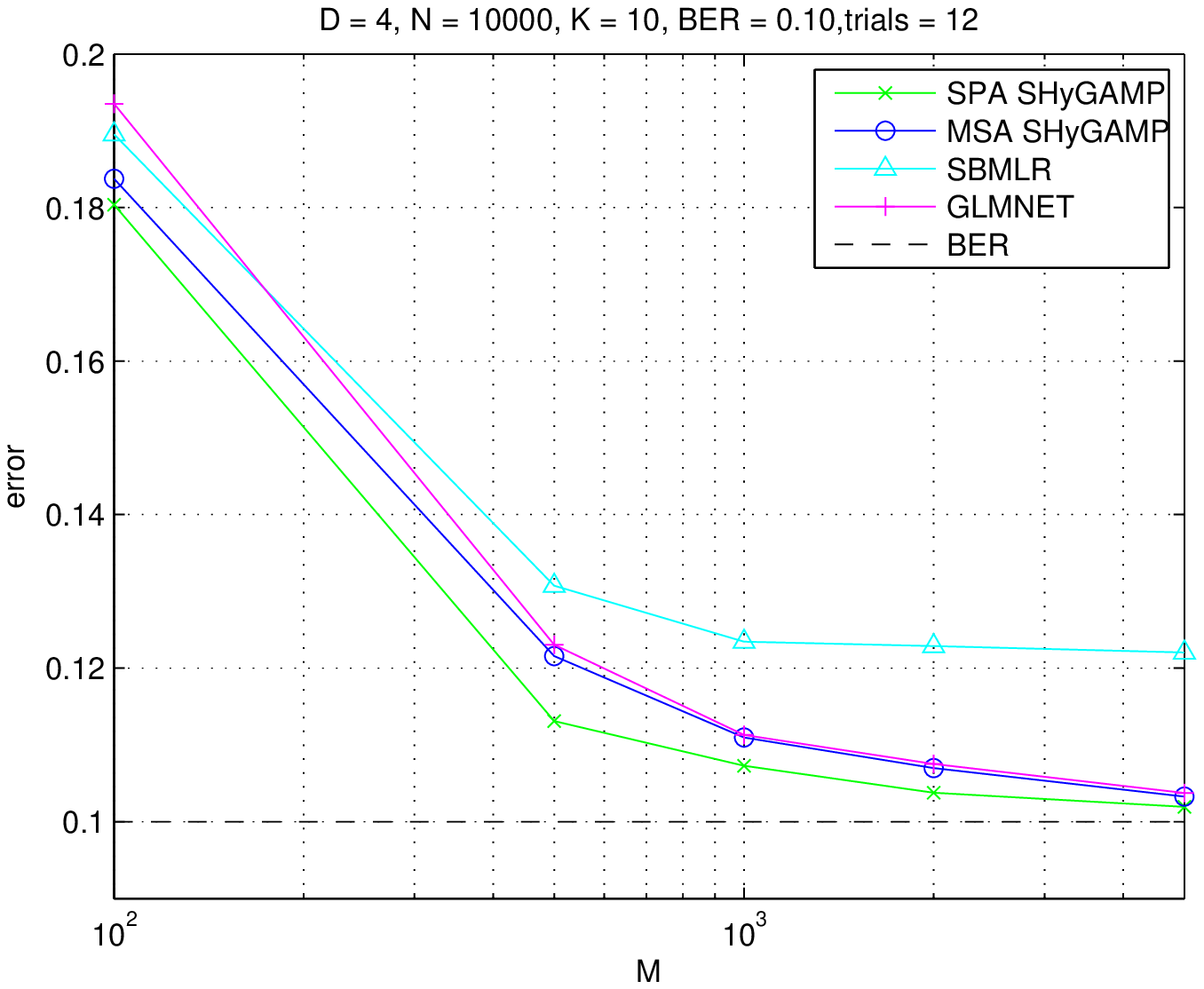}
\caption*{(a) Error}
\end{minipage}
\begin{minipage}{\figsize}
\psfrag{D = 4, N = 10000, K = 10, BER = 0.10,trials = 12}{}
\psfrag{time}[b][b][0.8]{\textsf{Runtime [sec]}}
\psfrag{M}[t][t][0.8]{\textsf{Number of Training Samples $M$}}
\includegraphics[width=1\textwidth]{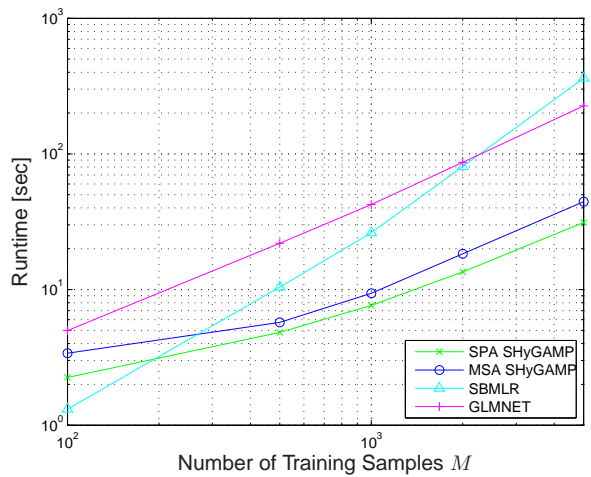}
\caption*{(b) Runtime}
\end{minipage}
\caption{Synthetic Experiment 1: \textb{expected} test-error rate and runtime versus $M$.  Here, $D = 4$, $N = 10000$, and $K = 10$.}
\label{fig:vsM}
\end{figure}

Figures~\ref{fig:vsM}a-b show the \textb{expected} test-error rate and runtime, respectively, versus the number of training examples, $M$, averaged over $12$ independent trials.
\Figref{vsM}a shows that, at all tested values of $M$, 
SPA-SHyGAMP gave the best error-rates 
and MSA-SHyGAMP gave the second best error-rates, although those reached by GLMNET were similar at large $M$.
Moreover, the error-rates of SPA-SHyGAMP, MSA-SHyGAMP, and GLMNET all converged towards the BER as $M$ increased, whereas that of SBMLR did not.
Since MSA-SHyGAMP, GLMNET, and SBMLR all solve the same $\ell_1$-regularized MLR problem, the difference in their error-rates can be attributed to the difference in their tuning of the regularization parameter $\lambda$.
\Figref{vsM}b shows that, 
for $M>500$, SPA-SHyGAMP was the fastest, followed by MSA-SHyGAMP, SBMLR, and GLMNET. 
Note that the runtimes of SPA-SHyGAMP, MSA-SHyGAMP, and GLMNET increased linearly with $M$, whereas the runtime of SBMLR increased quadratically with $M$. 

\begin{figure}
\captionsetup{font=\mycaptionfont}
\centering
\begin{minipage}{\figsize}
\psfrag{D = 4, M = 300, N = 30000, BER = 0.10,trials = 12}{}
\psfrag{error}[b][b][0.8]{\textsf{Test Error Rate}}
\psfrag{K}[t][t][0.8]{\textsf{True Sparsity $K$}}
\includegraphics[width=1\textwidth]{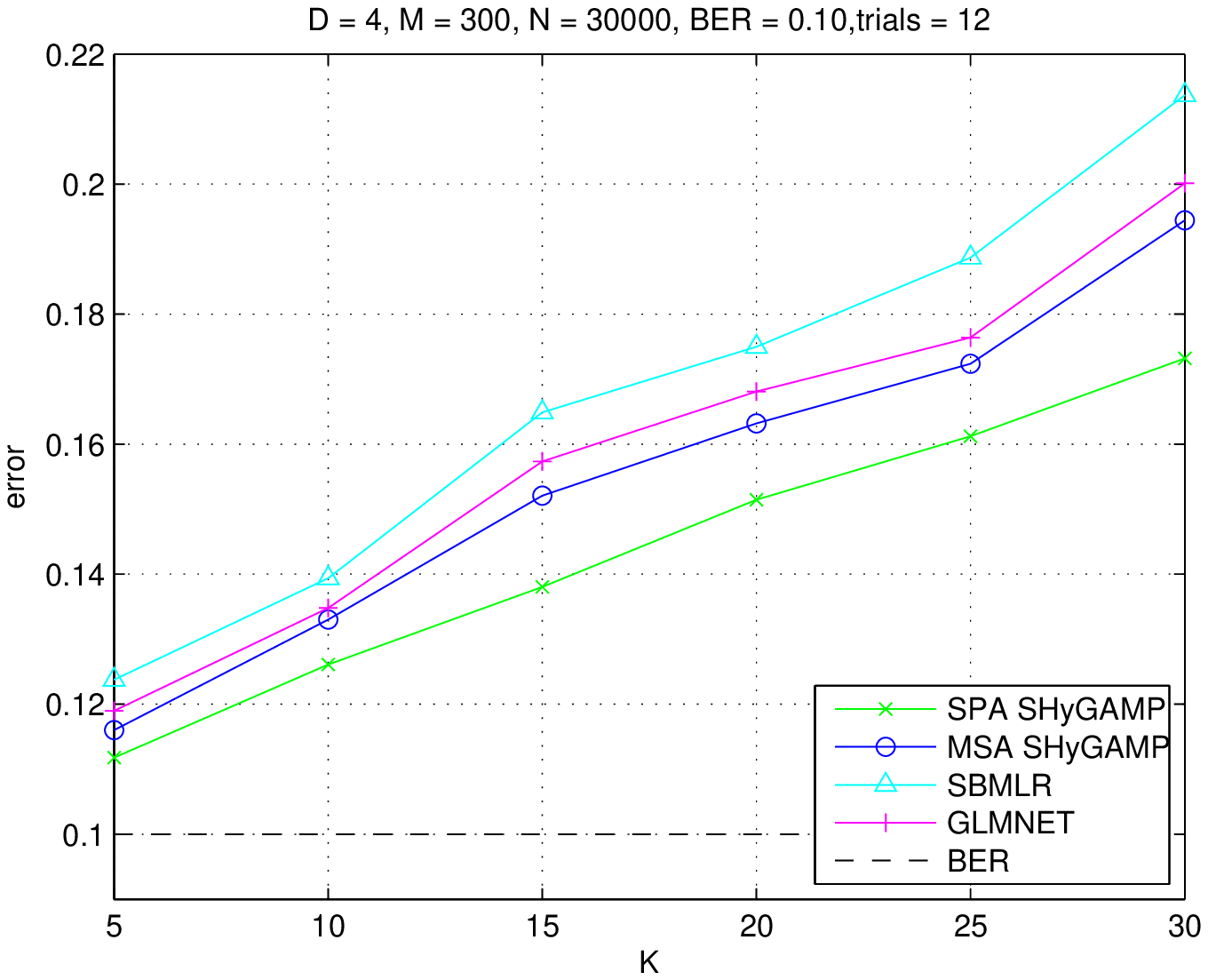}
\end{minipage}
\caption*{(a) Error}
\begin{minipage}{\figsize}
\psfrag{D = 4, M = 300, N = 30000, BER = 0.10,trials = 12}{}
\psfrag{time}[b][b][0.8]{\textsf{Runtime [sec]}}
\psfrag{K}[t][t][0.8]{\textsf{True Sparsity $K$}}
\includegraphics[width=1\textwidth]{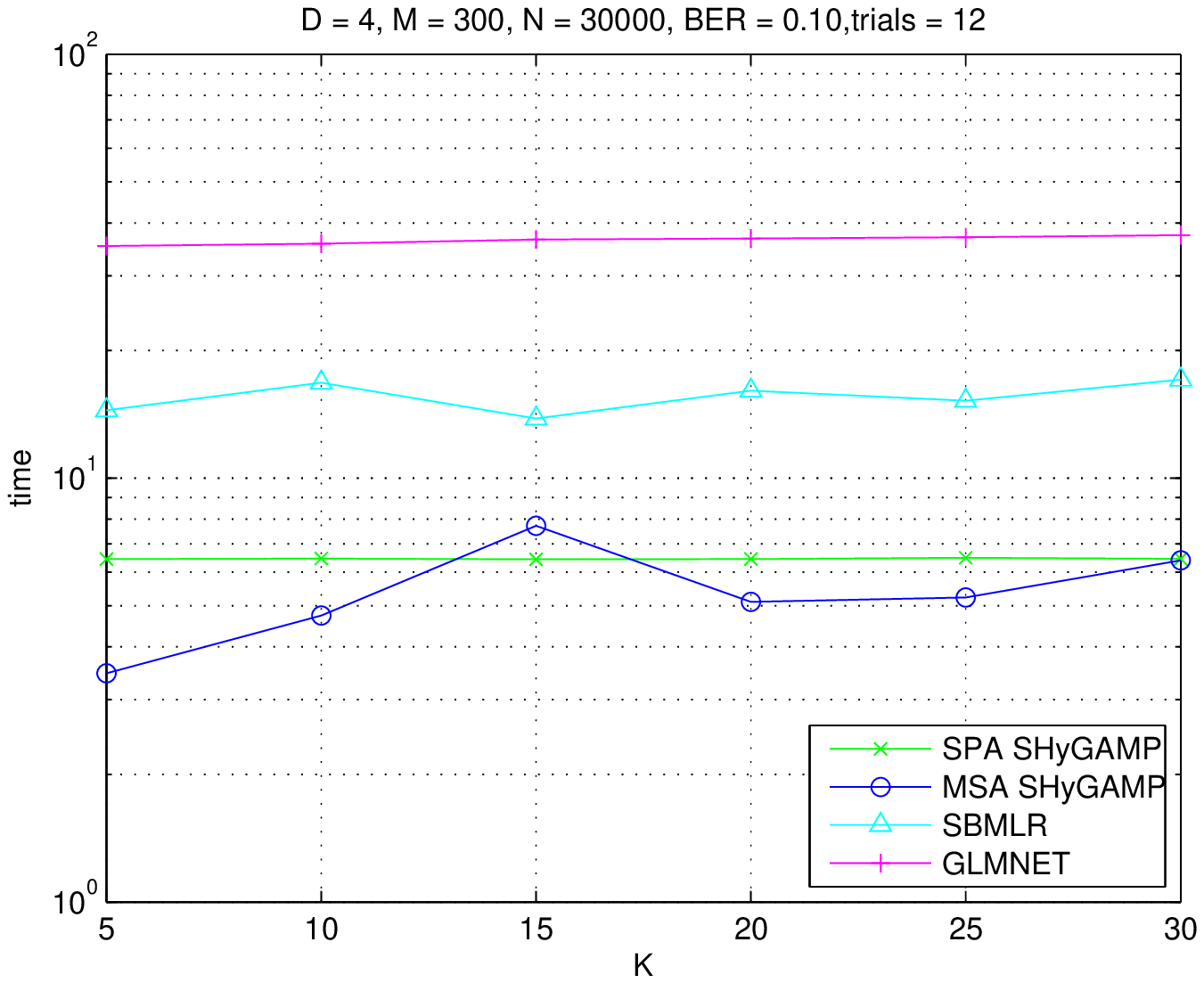}
\caption*{(b) Runtime}
\end{minipage}
\caption{Synthetic Experiment 2: \textb{expected} test-error rate and runtime versus $K$.  Here, $D = 4$, $M = 300$, and $N = 30000$.}
\label{fig:vsK}
\end{figure}

Figures~\ref{fig:vsK}a-b show the \textb{expected} test-error rate and runtime, respectively, versus feature-vector sparsity, $K$, averaged over $12$ independent trials.
\Figref{vsK}a shows that, at all tested values of $K$,
SPA-SHyGAMP gave the best error-rates  
and MSA-SHyGAMP gave the second best error-rates.
\Figref{vsK}b shows that 
SPA-SHyGAMP and MSA-SHyGAMP gave the fastest runtimes.
All runtimes were approximately invariant to $K$. 
 
\begin{figure}
\captionsetup{font=\mycaptionfont}
\centering
\begin{minipage}{\figsize}
\psfrag{D = 4, M = 200, K = 10, BER = 0.10,trials = 12}{}
\psfrag{error}[b][b][0.8]{\textsf{Test Error Rate}}
\psfrag{N}[t][t][0.8]{\textsf{Number of Features $N$}}
\includegraphics[width=1\textwidth]{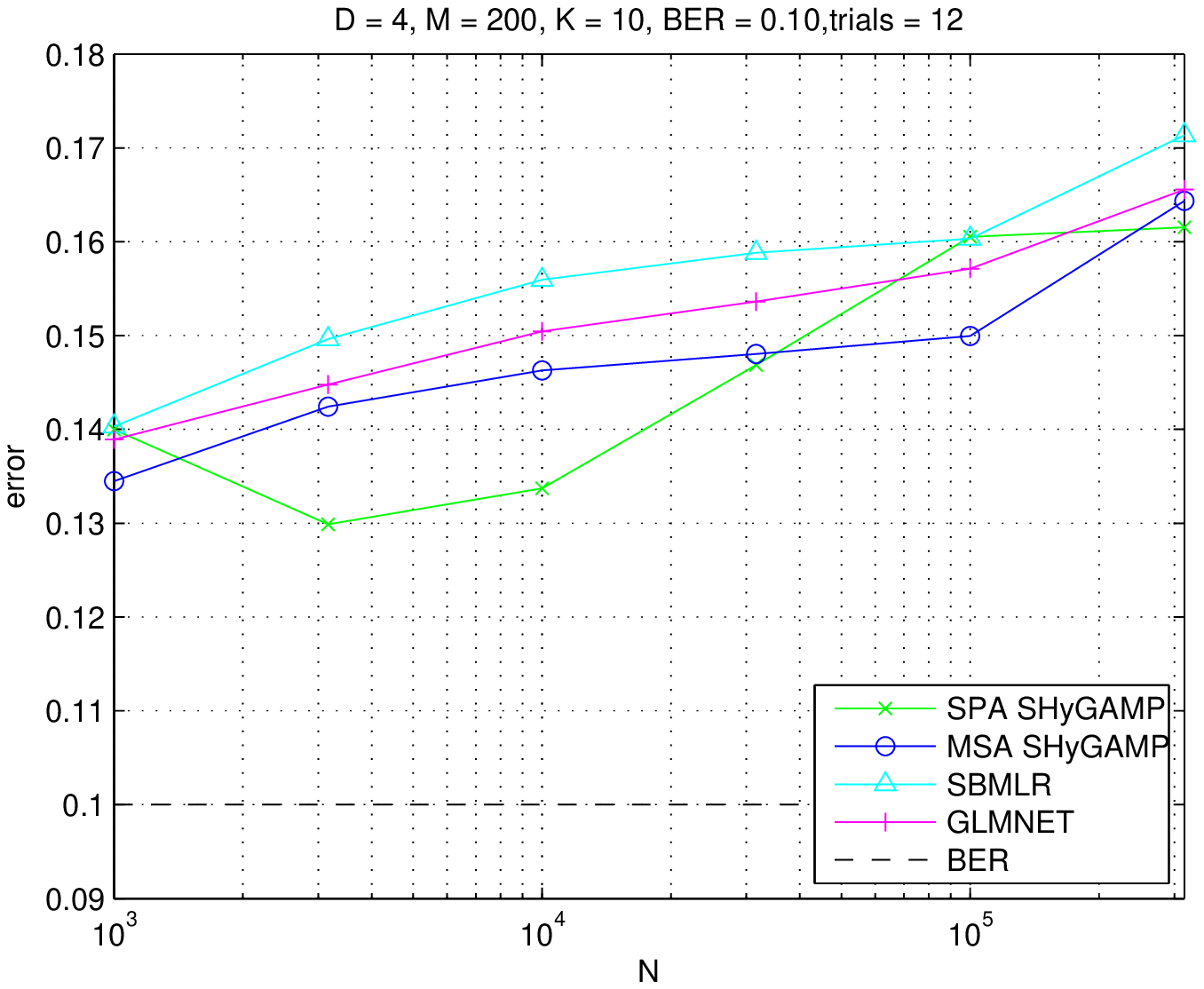}
\caption*{(a) Error}
\end{minipage}
\begin{minipage}{\figsize}
\psfrag{D = 4, M = 200, K = 10, BER = 0.10,trials = 12}{}
\psfrag{time}[b][b][0.8]{\textsf{Runtime [sec]}}
\psfrag{N}[t][t][0.8]{\textsf{Number of Features $N$}}
\includegraphics[width=1\textwidth]{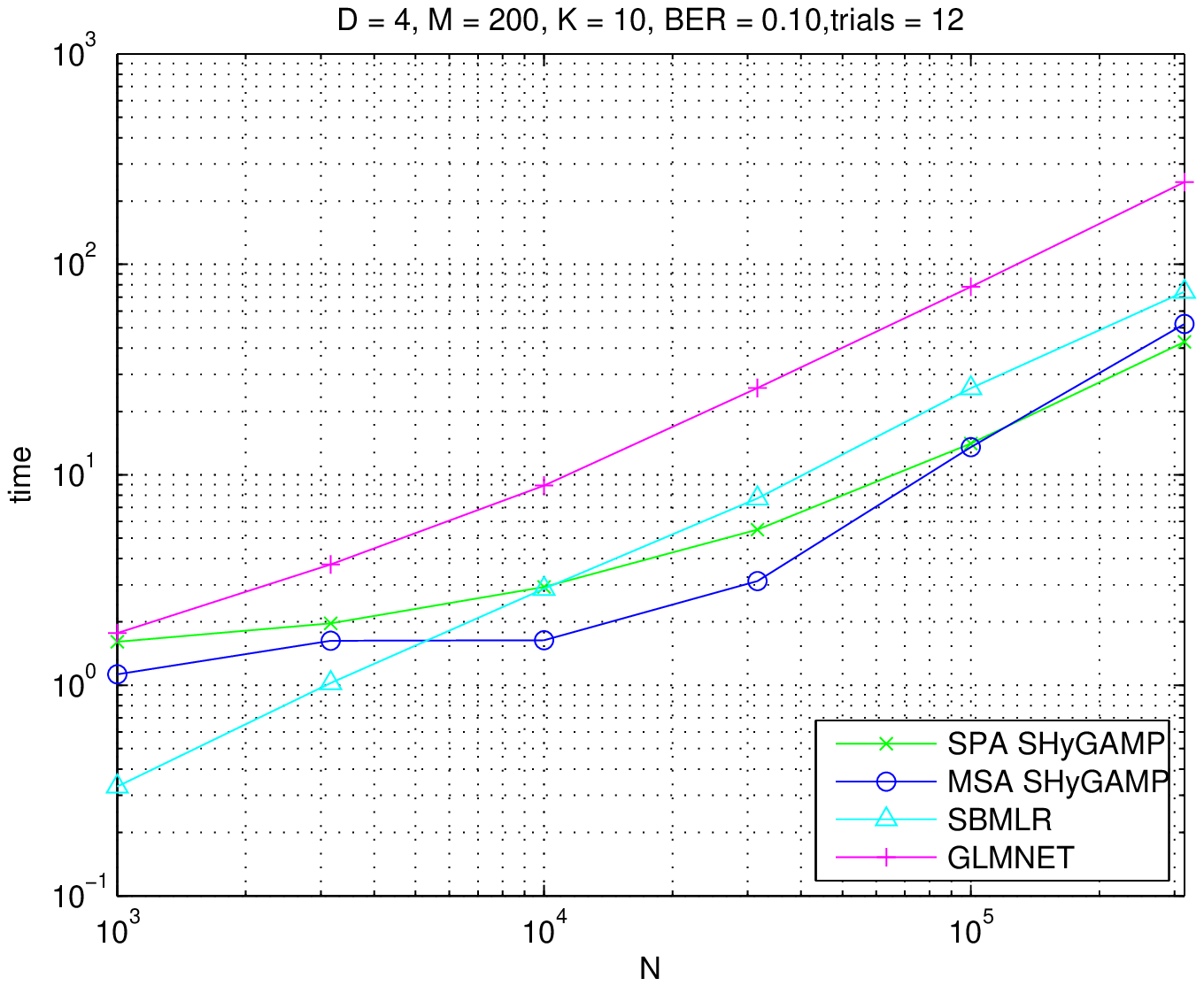}
\caption*{(b) Runtime}
\end{minipage}
\caption{Synthetic Experiment 3: \textb{expected} test-error rate and runtime versus $N$.  Here, $D = 4$, $M = 200$, and $K = 10$.}
\label{fig:vsN}
\end{figure}

Figures~\ref{fig:vsN}a-b show the \textb{expected} test-error rate and runtime, respectively, versus the number of features, $N$, averaged over $12$ independent trials.
\Figref{vsN}a shows that, at all tested values of $N$,
MSA-SHyGAMP gave lower error-rates than SBMLR and GLMNET.
Meanwhile, SPA-SHyGAMP gave the lowest error-rates for certain values of $N$. 
\Figref{vsN}b shows that
SPA-SHyGAMP and MSA-SHyGAMP gave the fastest runtimes for $N\geq 10000$, 
while SBMLR gave the fastest runtimes for $N\leq 3000$.
All runtimes increased linearly with $N$.
 

\subsection{Example of SURE tuning} \label{sec:sims_sure}

\begin{figure}[t]
\captionsetup{font=\mycaptionfont}
\centering
\psfrag{N = 30000, M = 300, D = 4, K = 25, BER = 0.100, trials = 12}{}
\psfrag{error}[b][b][0.8]{\textsf{Test Error Rate}}
\psfrag{lambda}[t][t][0.8]{\textsf{Regularization Parameter $\lambda$}}
\includegraphics[width=\figsize]{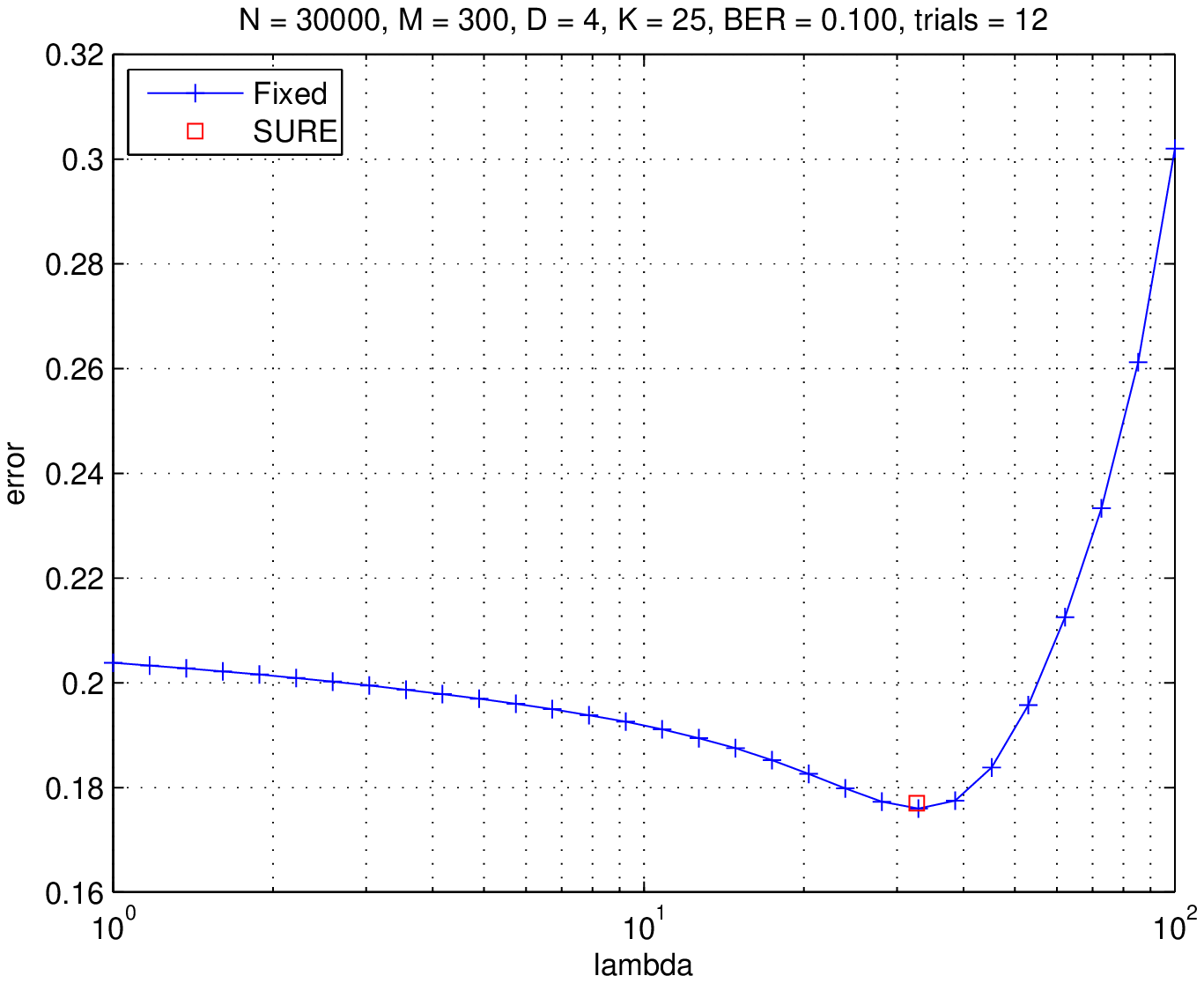}
\caption{\textb{Synthetic experiment 4: expected test-error rate versus regularization parameter $\lambda$ for fixed-$\lambda$ MSA-SHyGAMP.  Here, $D=4$, $M=300$, $N=30000$, and $K=25$.}
Also shown is the average test-error rate for SURE-tuned MSA-SHyGAMP plotted at the average value of $\hat{\lambda}$.}
\label{fig:sure}
\end{figure}

Although the good error-rate performance of MSA-SHyGAMP in \secref{sims_synth} suggests that the SURE $\lambda$-tuning method from \secref{param_tune_SURE} is working reliably, we now describe a more direct test of its behavior.
Using synthetic data generated as described in \secref{sims_synth} with $D=4$ classes, $N = 30000$ features, $M = 300$ examples, and sparsity $K = 25$, we ran MSA-SHyGAMP using various fixed values of $\lambda$.
\textb{In the sequel, we refer to this experiment as ``Synthetic Experiment 4.''}
The resulting \textb{expected} test-error rate versus $\lambda$ (averaged over $10$ independent realizations) is shown in \figref{sure}.
For the same realizations, we ran MSE-SHyGAMP with SURE-tuning and plot the resulting error-rate and average $\hat{\lambda}$ in \figref{sure}.
From \figref{sure}, we see that the SURE $\lambda$-tuning method matched both the minimizer and the minimum of the error-versus-$\lambda$ trace of fixed-$\lambda$ MSA-SHyGAMP.

\subsection{Micro-array gene expression} \label{sec:sims_gene}

Next we consider classification and feature-selection using micro-array gene expression data.
Here, the labels indicate which type of disease is present (or no disease) and the features represent gene expression levels.
The objective is i) to determine which subset of genes best predicts the various diseases and ii) to classify whether an (undiagnosed) patient is at risk for any of these diseases based on their gene profile.

We tried two datasets: one from Sun et al.\ \cite{Sun:CAN:06} and one from Bhattacharjee et al.\ \cite{Bhattacharjee:PNAS:01}.
The Sun dataset includes $M=179$ examples, $N=54613$ features, and $D=4$ classes; and
the Bhattacharjee dataset includes $M = 203$ examples, $N=12600$ features, and $D=5$ classes.
With the Sun dataset, we applied a $\log_2(\cdot)$ transformation and z-scored prior to processing, while with Bhattacharjee we simply z-scored (since the dataset included negative values).

\color{\blue}
The test-error rate was estimated as follows for each dataset.
We consider a total of $T$ ``trials.'' 
For the $t$th trial, we 
i) partition the dataset into a training subset of size $\Mtraint$ and a test subset of size $\Mtestt$, 
ii) design the classifier using the training subset, and
iii) apply the classifier to the test subset, recording the test errors $\{e_{tm}\}_{m=1}^{\Mtestt}$, where $e_{tm}\in\{0,1\}$ indicates whether the $m$th example was in error.
We then estimate the average test-error rate using the empirical average $\hat{\mu}\defn \Mtest^{-1} \sum_{t=1}^T \sum_{m=1}^{\Mtestt} e_{tm}$, where $\Mtest=\sum_{t=1}^T\Mtestt$.
If the test sets are constructed without overlap, we can model $\{e_{tm}\}$ as i.i.d.\ $\text{Bernoulli}(\mu)$, where $\mu$ denotes the true test-error rate.
Then, since $\hat{\mu}$ is $\text{Binomial}(\mu,\Mtest)$, the standard deviation (SD) of our error-rate estimate $\hat{\mu}$ is $\sqrt{\var\{\hat{\mu}\}}=\sqrt{\mu(1-\mu)/\Mtest}$.
Since $\mu$ is unknown, we approximate the SD by $\sqrt{\hat{\mu}(1-\hat{\mu})/\Mtest}$.
\color{black}


Tables~\ref{tab:bio} and \ref{tab:lung} show, for each algorithm, the test-error rate \textb{estimate $\hat{\mu}$, the approximate SD $\sqrt{\hat{\mu}(1-\hat{\mu})/\Mtest}$ of the estimate,} the average runtime, and two metrics for the sparsity of $\hvec{X}$.
The $\|\hvec{X}\|_0$ metric quantifies the number of non-zero entries in $\hvec{X}$ (i.e., absolute sparsity), while the $\hat{K}_{99}$ metric quantifies the number of entries of $\hvec{X}$ needed to reach $99\%$ of the Frobenius norm of $\hvec{X}$ (i.e., effective sparsity).
\textb{We note that the reported values of $\hat{K}_{99}$ and $\|\hvec{X}\|_0$ represent the average over the $T$ folds.}
\textb{For both the Sun and Bhattacharjee datasets, we used $T=19$ trials and $\Mtestt=\floor{M/20}~\forall t$.}

\tabref{bio} shows results for the Sun dataset. 
There we see that MSA-SHyGAMP gave the best test-error rate, although the other algorithms were not far behind \textb{and all error-rate estimates were within the estimator standard deviation.}
SPA-SHyGAMP was the fastest algorithm and MSA-SHyGAMP was the second fastest, with the remaining algorithms running 
\color{\blue}
$3\times$ to $5\times$ slower than SPA-SHyGAMP.
GLMNET's weights were the sparsest according to both sparsity metrics. 
SPA-SHyGAMP's weights had the second lowest value of $\hat{K}_{99}$, even though they were technically non-sparse (i.e., $\|\hvec{X}\|_0=218\,452=ND$) as expected.
Meanwhile, MSA-SHyGAMP's weights were the least sparse according to the $\hat{K}_{99}$ metric.
\color{black}

\tabref{lung} shows results for the Bhattacharjee dataset.
\color{\blue}
In this experiment, SPA-SHyGAMP and SBMLR were tied for the best error rate, MSA-SHyGAMP was $0.5$ standard-deviations worse, and GLMNET was $1.2$ standard-deviations worse.
However, SPA-SHyGAMP ran about twice as fast as SBMLR, and $4\times$ as fast as GLMNET.
As in the Sun dataset, SPA-SHyGAMP returned the sparsest weight matrix according to the $\hat{K}_{99}$ metric.
The sparsities of the weight matrices returned by the other three algorithms were similar to one another in both metrics.
Unlike in the Sun dataset, MSA-SHyGAMP and SBMLR had similar runtimes (which is consistent with \figref{vsN}b since $N$ is lower here than in the Sun dataset). 
\color{black}


\begin{table}[t]
\centering
\footnotesize
\color{\blue}
\begin{tabular}{|l|c|c|c|c|}
\hline 
Algorithm & \% Error (SD) & Runtime (s) & $\hat{K}_{99}$ & $\|\hvec{X}\|_0$ \\[2pt]  \hline 
SPA-SHyGAMP & 33.3 (3.8)  &  $\phantom{1}$6.86  &   20.05  & 218\,452 \\ 
MSA-SHyGAMP & 31.0 (3.7)  &  13.59  &  93.00  & 145.32  \\ 
SBMLR       & 31.6 (3.7)  &  22.48  &   49.89  & 72.89   \\
GLMNET      & 33.9 (3.8)  &  31.93  &   10.89  & 16.84   \\ \hline
\end{tabular}
\captionsetup{font=\mycaptionfont}
\caption{\textb{Estimated} test-error rate, standard deviation \textb{of estimate}, runtime, and sparsities for the Sun dataset.}
\label{tab:bio}
\end{table}


\begin{table}[t]
\centering
\footnotesize
\color{\blue}
\begin{tabular}{|l|c|c|c|c|}
\hline 
Algorithm & \% Error (SD) & Runtime (s) & $\hat{K}_{99}$ & $\|\hvec{X}\|_0$ \\[2pt]  \hline 
SPA-SHyGAMP &  $\phantom{1}$9.5 (2.1) &  $\phantom{1}$3.26 & 16.15 & 63\,000   \\ 
MSA-SHyGAMP & 10.5 (2.2) &  $\phantom{1}$6.11 & 55.20 & 84.65 \\ 
SBMLR       &  $\phantom{1}$9.5 (2.1) &  $\phantom{1}$6.65 & 44.25 & 79.10   \\
GLMNET      & 12.0 (2.4) & 13.67 & 49.65 & 89.40   \\ \hline
\end{tabular}
\captionsetup{font=\mycaptionfont}
\caption{\textb{Estimated} test-error rate, standard deviation \textb{of estimate}, runtime, and sparsities for the Bhattacharjee dataset.}
\label{tab:lung}
\end{table}

\subsection{Text classification with the RCV1 dataset} \label{sec:sims_text}

Next we consider text classification using the Reuter's Corpus Volume 1 (RCV1) dataset \cite{Lewis:MLR:04}.
Here, each sample $(y_m,\vec{a}_m)$ represents a news article, where $y_m$ indicates the article's topic and $\vec{a}_m$ indicates the frequencies of common words in the article.
The version of the dataset that we used\footnote{\url{http://www.csie.ntu.edu.tw/~cjlin/libsvmtools/datasets/multiclass.html}} contained $N=47\,236$ features and $53$ topics.
However, we used only the first $D=25$ of these topics (to reduce the computational demand). 
Also, we retained the default training and test partitions, which resulted in the use of $M=14\,147$ samples for training and $469\,571$ samples for testing.

The RCV1 features are very sparse (only 0.326\% of the features are non-zero) and non-negative, which conflicts with the standard assumptions used for the derivation of AMP algorithms: that $\vec{A}$ is i.i.d.\ zero-mean and sub-Gaussian.
\color{\blue}
Interestingly, the RCV1 dataset also caused difficulties for SBMLR, which diverged under default settings.
This divergence was remedied by decreasing the value of a step-size parameter\footnote{See the variable \texttt{scale} on lines 129 and 143 of \texttt{sbmlr.m}.} to $0.1$ from the default value of $1$.
\color{black}

\begin{figure}[t]
\centering
\psfrag{error}[b][b][0.8]{\textsf{Test Error Rate}}
\psfrag{time}[t][t][0.8]{\textsf{Cumulative Runtime [sec]}}
\includegraphics[width=\figsize]{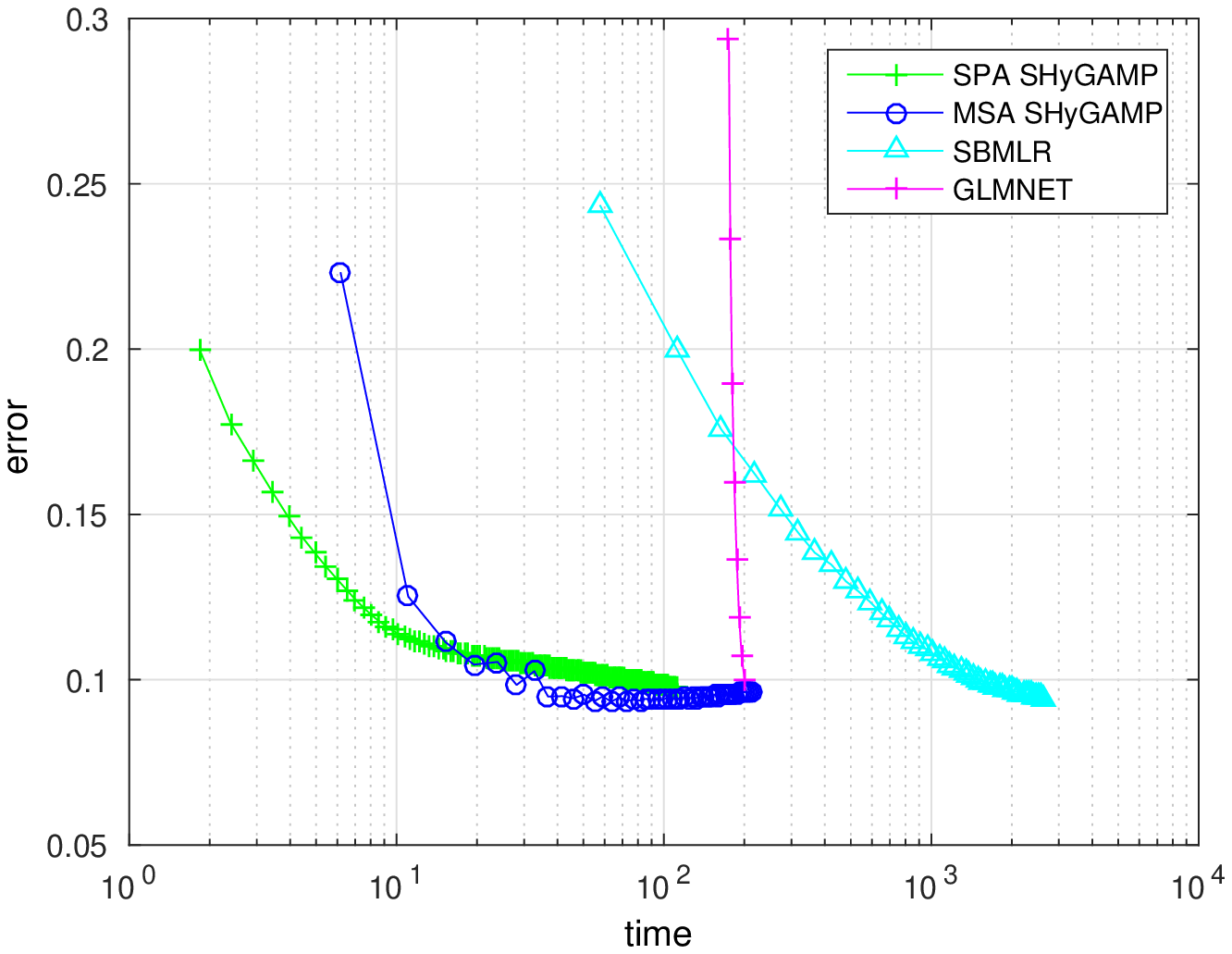}
\captionsetup{font=\mycaptionfont}
\caption{\textb{Test-error rate versus runtime for the RCV1 dataset.}}
\label{fig:rcv1_evt}
\end{figure}

\Figref{rcv1_evt} shows test-error rate versus runtime for SPA-SHyGAMP, MSA-SHyGAMP, SBMLR, \textb{and GLMNET} on the RCV1 dataset.
\textb{In the case of SPA-SHyGAMP, MSA-SHyGAMP and SBMLR}, each point in the figure represents one iteration of the corresponding algorithm.
\color{\blue}
For GLMNET, each data-point represents one iteration of the algorithm \emph{after} its cross-validation stage has completed.\footnote{%
\textb{GLMNET spent most of its time on cross-validation.
After cross-validation, GLMNET took $25.26$ seconds to run, which is similar to the total runtimes of SPA-SHyGAMP and MSE-SHyGAMP.}
} 
We used $2$ CV folds (rather than the default $10$) in this experiment to avoid excessively long runtimes.
\color{black}
The figure shows that the SHyGAMP algorithms converged more than an order-of-magnitude faster than SBMLR \textb{and GLMNET}, although the final error rates were similar.
SPA-SHyGAMP displayed faster initial convergence, but MSA-SHyGAMP eventually caught up.


\subsection{MNIST handwritten digit recognition} \label{sec:sims_digits}

Finally, we consider handwritten digit recognition using the Mixed National Institute of Standards and Technology (MNIST) dataset\cite{LeCun:PROC:98}. 
This dataset consists of $70\,000$ examples, where each example is an $N=784$ pixel image of one of $D=10$ digits between $0$ and $9$. 
\textb{These features were again non-negative, which conflicts with the standard AMP assumption of i.i.d.\ zero-mean $\vec{A}$.}

\begin{figure}[t]
\centering
\psfrag{t = 50 trials}{}
\psfrag{test error rate}[b][b][0.8]{\textsf{Test Error Rate}}
\psfrag{training samples}[t][t][0.8]{\textsf{Number of Training Samples $M$}}
\includegraphics[width=\figsize]{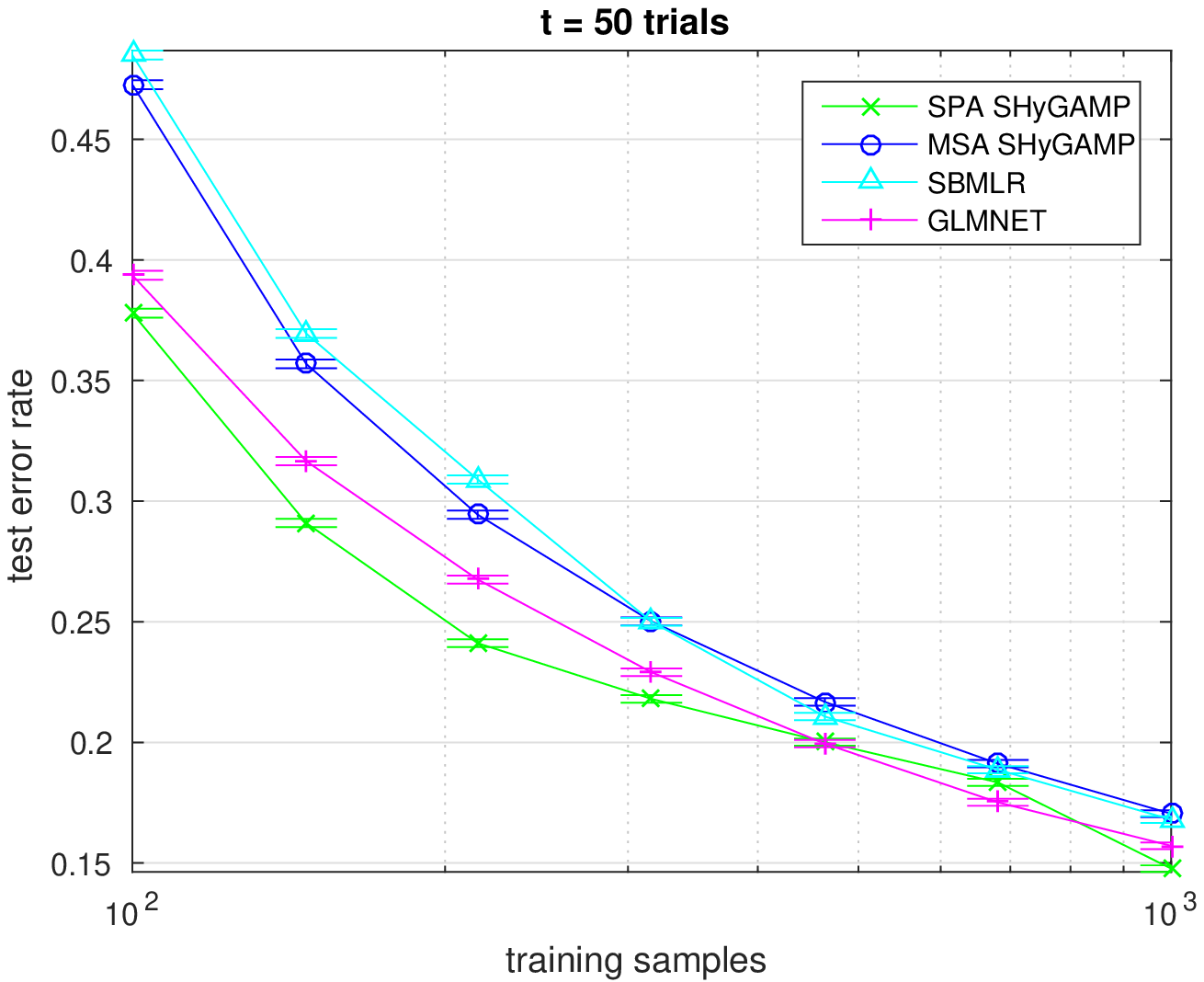}
\captionsetup{font=\mycaptionfont}
\caption{\textb{Estimated test-error rate versus $M$ for the MNIST dataset, with error-bars indicating the standard deviation of the estimate.}}
\label{fig:mnist}
\end{figure}

Our experiment characterized test-error rate versus the number of training examples, $M$, for the SPA-SHyGAMP, MSA-SHyGAMP, SBMLR, \textb{and GLMNET} algorithms.
For each value of $M$, we performed $50$ Monte-Carlo trials.
\color{\blue}
In each trial, $M$ training samples were selected uniformly at random and the remainder of the data were used for testing.
\Figref{mnist} shows the average estimated test-error rate $\hat{\mu}$ versus the number of training samples, $M$, for the algorithms under test.
The error-bars in the figure correspond to the average of the per-trial estimated SD over the $50$ trials.
For SBMLR, we reduced the stepsize to $0.5$ from the default value of $1$ to prevent a significant degradation of test-error rate.
The figure shows SPA-SHyGAMP attaining significantly better error-rates than the other algorithms at small values of $M$ (and again at the largest value of $M$ considered for the plot). 
For this plot, $M$ was chosen to focus on the $M<N$ regime.
\color{black}


\color{black}

\section{Conclusion}		\label{sec:conc}

For the problem of multi-class linear classification and feature selection, we proposed several AMP-based approaches to sparse multinomial logistic regression.
We started by proposing two algorithms based on HyGAMP \cite{Rangan:ISIT:12},
one of which finds the maximum a posteriori (MAP) linear classifier based on the multinomial logistic likelihood and a Laplacian prior, and the other of which finds an approximation of the test-error-rate minimizing linear classifier based on the multinomial logistic likelihood and a Bernoulli-Gaussian prior.
The numerical implementation of these algorithms is challenged, however, by the need to solve $D$-dimensional inference problems of multiplicity $M$ at each HyGAMP iteration.
Thus, we proposed simplified HyGAMP (SHyGAMP) approximations based on a diagonalization of the message covariances and a careful treatment of the $D$-dimensional inference problems.
In addition, we described EM- and SURE-based methods to tune the hyperparameters of the assumed statistical model. 
Finally, using both synthetic and real-world datasets, we demonstrated improved error-rate and runtime performance relative to the state-of-the-art SBMLR \cite{Friedman:JSS:10} and GLMNET \cite{Cawley:NIPS:07} \textb{approaches to sparse MLR}.  



\bibliographystyle{ieeetr}
\bibliography{macros_abbrev,stc,books,misc,sparse,machine,gamp_classification}


\def\baselinestretch{1.0}



\end{document}